\documentclass[12pt,a4paper]{article}
\usepackage{epsfig}

\begin{document}
\begin{flushright}AEI-2011-048
\end{flushright}
\begin{center}

\vspace{1cm}

{\bf \large ON MHV FORM FACTORS IN SUPERSPACE FOR $\mathcal{N}=4$
SYM THEORY} \vspace{2cm}

{\bf \large L. V. Bork$^{2}$, D. I. Kazakov$^{1,2}$, G. S.
Vartanov$^{3}$}\vspace{0.5cm}

{\it $^1$Bogoliubov Laboratory of Theoretical Physics, Joint
Institute for Nuclear Research, Dubna, Russia, \\
$^2$Institute for Theoretical and Experimental Physics, Moscow, Russia, \\
$^3$Max-Planck-Institut f\"ur Gravitationsphysik,
Albert-Einstein-Institut 14476 Golm, Germany.}\vspace{1cm}

\abstract{In this paper we develop a supersymmetric version of a
unitarity cut method for form factors of operators from the
$\mathcal{N}=4$ stress-tensor current supermultiplet $T^{AB}$. The
relation between the super form factor with super momentum equal to
zero and the logarithmic derivative of the superamplitude with
respect to the coupling constant is discussed and verified at the
tree- and one-loop level for any MHV $n$-point ($n \geq 4$) super
form factor. The explicit $\mathcal{N}=4$ covariant expressions for
n-point MHV tree- and one-loop form factors are obtained. As well,
the ansatz for the two-loop three-point MHV super form factor is
suggested in the planar limit, based on the reduction procedure for
the scalar integrals suggested in our previous work. The different
soft and collinear limits in the MHV sector at the tree- and
one-loop level are discussed.}
\end{center}


%
%
\newpage
\tableofcontents{}\vspace{0.5cm}
%

\renewcommand{\theequation}{\thesection.\arabic{equation}}
\section{Introduction}\label{s1}
Much attention in the past few years has been  paid to the study of
the planar limit for the scattering amplitudes in the
$\mathcal{N}=4$ SYM theory. It is believed that the hidden
symmetries responsible for integrability properties of this theory
completely fix the structure of the amplitudes (the $S$-matrix of
the theory). The hints that the $S$-matrix for the $\mathcal{N}=4$
SYM theory can be fixed by some underlying integrable structure were
found at weak
\cite{BeisertYangianRev,BeisertYangianAmpl,Gorsky:2009ew} and strong
\cite{Gorsky:2009ew,Alday:2010vh} coupling regimes.

The progress in understanding of the $S$-matrix structure of the
$\mathcal{N}=4$ SYM theory has been made mainly due to achievements
in the development of the computational methods beyond the ordinary
textbook Feynman rules, such as MHV vertex expansion \cite{rec0},
recursion relations \cite{rec1}, unitarity cut constructibility
techniques \cite{Bern1}, generalized unitarity \cite{Britto:2004nc}
and their $\mathcal{N}=4$ covariant generalizations
\cite{N=4onshellmethods1,N=4onshellmethods2}.

There is another class of objects of interest in the $\mathcal{N}=4$ SYM
theory which resemble the amplitudes -- the form factors which are
the matrix elements of the form
\begin{equation}
\langle0|\mathcal{O}|p_1^{\lambda_1}, \ldots,
p_n^{\lambda_n}\rangle,
\end{equation}
where $\mathcal{O}$ is some gauge invariant operator which acts on
the vacuum and produces some state $|p_1^{\lambda_1}, \ldots,
p_n^{\lambda_n} \rangle$ with momenta $p_1, \ldots, p_n$ and
helicities $\lambda_1, \ldots, \lambda_n$\footnote{Note that
scattering amplitudes in ''all ingoing" notation can schematically
be written as $\langle 0|p_1^{\lambda_1}, \ldots, p_n^{\lambda_n}
\rangle$. }. The $S$-matrix operator is assumed in both cases.

The two-point form factor was studied long time ago in
\cite{vanNeerven:1985ja} and later, using the $\mathcal{N}=3$
superfield formalism, the tree-level form factors were derived in
\cite{Perturbiner}. Recently, the strong coupling limit of  form
factors has been studied in \cite{Maldacena:2010kp} and the weak
coupling regime in \cite{FormFMHV,BKV_FormFN=1}. The motivations for
the systematic study of these objects are
\begin{itemize}
\item it might help in understanding of
the symmetry properties of the amplitudes
\cite{BeisertYangianRev,BeisertYangianAmpl}. It is believed that the
symmetries completely fix the amplitudes of the $\mathcal{N}=4$ SYM
theory and  it is interesting to see whether they fix/restrict the form factors as well;

\item the form factors are the intermediate objects between
the fully on-shell quantities such as the amplitudes and the fully
off-shell quantities such as the correlation functions (which are
one of the central objects in AdS/CFT). Since the powerful
computational methods have appeared recently for the amplitudes in
$\mathcal{N}=4$ SYM \cite{N=4onshellmethods1,N=4onshellmethods2}, it
would be desirable to have some analog  of them for the correlation
functions \cite{Raju:2011mp}. The understanding of the structure of
form factors and the development of computational methods  might
shed light on the correlation functions;

\item recently, it was observed \cite{Eden:2010zz} that in the limit when
the distances between the operators in the correlation functions
become light-like, there is a simple relation between the
$n+l$-point correlation function with $l$ Lagrangian insertions and
the integrand of the MHV $n$-point amplitude at the $l$ loop level.
The form factors, as was explained above, are the intermediate objects
between completely on-shell/off-shell quantities, so they might help
in understanding the relations discussed in \cite{Eden:2010zz};

\item also, it might be useful for understanding of
the relation between the conventional description of the gauge
theory in terms of local operators and its (possible) description in
terms of Wilson loops. The latter fact is the so-called
amplitude/Wilson loop duality which originated in
\cite{minSurface4point,dualKorch,dual}. This duality was intensively
studied in the weak and strong coupling regimes and tested in
different cases, and its generalizations to the non-MHV amplitudes
were proposed in \cite{Huot}. Recently, a similar relation between
certain form factors and Wilson loops was observed at the one-loop
level in the weak coupling regime \cite{FormFMHV} which should be
verified at a higher loop level.
\end{itemize}

To make progress in the above-mentioned directions, the perturbative
computations at several first orders of perturbative theory (PT) are
likely required. For this purpose, proper computational tools beyond
the ordinary textbook Feynman rules which are bulky at higher order
of PT and/or at a large number of external particles are desirable.
The recent attempts at systematic study of form factors using the
$\mathcal{N}=1$ superfield formalism has been carried out in
\cite{BKV_FormFN=1} and the component version of recursion relations
and unitarity cut constructibility techniques have been applied in
\cite{FormFMHV}. The latter method is likely more efficient due to
the experience with the amplitude computations but still  there is a
lack of full $\mathcal{N}=4$ supersymmetry covariance, which makes
the summation over the intermediate states at a higher loop level
($l\geq2$) rather cumbersome \cite{SuperSums}.

The aim of this paper is to discuss the $\mathcal{N}=4$
generalization of the unitarity cut constructibility techniques for
the form factors of  the $\mathcal{N}=4$ stress tensor current
supermultiplet $T^{AB}$ (more accurately, its chiral truncation). We
introduce the "superstate super form factor" at the tree level and
explicitly define its MHV part. Then we re-derive the one-loop
correction to its MHV part for an arbitrary number of legs in
external state $|p_1^{\lambda_1}, \ldots, p_n^{\lambda_n}\rangle$ in
the covariant $\mathcal{N}=4$ notation. We also suggest an ansatz
for the 3-point MHV super form factor at two loops based on the
conjecture that the basis of scalar integrals for form factors can
be obtained from that of dual pseudoconformal integrals arising in
calculations of the amplitudes via the reduction procedure suggested
in \cite{BKV_FormFN=1}.  At the end, we also discuss various soft
and collinear limits at the tree- and one-loop level.

\section{Form factors of the $\mathcal{N}=4$ stress
tensor  supermultiplet and half-BPS operators in superspace}

\subsection{Preliminaries and notation}
It is convenient to describe the pure on-shell scattering amplitudes
using the so-called  $\mathcal{N}=4$ on-shell momentum superspace
\cite{DualConfInvForAmplitudesCorch}. This superspace is
parameterized in terms of $SL(2,C)$ spinors $\lambda_{\alpha},
\tilde{\lambda}_{\dot{\alpha}}, \alpha, \dot{\alpha}=1,2$ and
Grassmannian coordinates $\eta^A, A=1, \ldots,4$ which are Lorentz
scalars and $SU(4)_R$ vectors
\begin{equation}
\mbox{On-shell $\mathcal{N}=4$ momentum superspace} =
\{\lambda_{\alpha}, \tilde{\lambda}_{\dot{\alpha}}, \eta^A\}.
\end{equation}
Note that this superspace is chiral (i.e. it can be parameterized
only in terms of  Grassmannian variables $\eta$, or, equivalently,
 in terms of their conjugated partners $\bar{\eta}$). The
 generators of supersymmetry algebra relevant for our discussion,
in the case when we want to describe the n-particle amplitude, are
realized in on-shell momentum superspace as
\begin{eqnarray}\label{SUSYGenLightCone}
\mbox{4
translations } p_{\alpha\dot{\alpha}} &=& \sum_{i=1}^n\lambda_{\alpha}^i\tilde{\lambda}_{\dot{\alpha}}^i,\nonumber\\
\mbox{8 supercharges } q_{\alpha}^A &=& \sum_{i=1}^n\lambda_{\alpha}^i\eta^A_i,\nonumber\\
\mbox{8 conjugated supercharges }
\bar{q}_{A\dot{\alpha}}&=&\sum_{i=1}^n\tilde{\lambda}_{\dot{\alpha}}^i\frac{\partial}{\partial\eta^A_i}
=\sum_{i=1}^n\tilde{\lambda}_{\dot{\alpha}}^i\partial_{iA},
\end{eqnarray}
where $\lambda_{\alpha}^i$ and $\tilde{\lambda}_{\dot{\alpha}}^i$
correspond to the i-th particle with on-shell momentum
$p^{i}_{\mu}(\sigma^{\mu})_{\alpha\dot{\alpha}}=
\lambda_{\alpha}^{i}\tilde{\lambda}_{\dot{\alpha}}^{i}$, $p_i^2=0$,
$(\sigma^{\mu})_{\alpha\dot{\alpha}}$ is the Pauli sigma matrix,
$\mu$ is the index of the Lorentz group vector representation. In
the on-shell momentum superspace the creation/annihilation operators
$$
\{g^-,~\Gamma^A,~\phi^{AB},~\bar{\Gamma}^A,~g^+\},
$$
of the $\mathcal{N}=4$ supermultiplet, for the on-shell states which are
two physical polarizations of gluons $|g^-\rangle, |g^+\rangle$,
four fermions $|\Gamma^A\rangle$ with positive and four fermions
$|\bar{\Gamma}^A\rangle$ with negative helicity, and three complex
scalars $|\phi^{AB}\rangle$ (anti-symmetric in $SU(4)_R$ indices
$AB$ ) can be combined together into one $\mathcal{N}=4$ invariant
superstate ("superwave-function")
$|\Omega_{i}\rangle$
\begin{eqnarray}\label{superstate}
|\Omega_{i}\rangle = \left(g^+_i + \eta^A\Gamma_{i,A} +
\frac{1}{2!}\eta^A\eta^B \phi_{i,AB} +
\frac{1}{3!}\eta^A\eta^B\eta^C \varepsilon_{ABCD}\bar{\Gamma}^D_i +
\frac{1}{4!}\eta^A\eta^B\eta^C\eta^D \varepsilon_{ABCD}g^-_i\right) |0\rangle,\nonumber\\
\end{eqnarray}
here $i$ corresponds to the on-shell momentum
$p^i_{\alpha\dot{\alpha}} =
\lambda_{\alpha}^i\tilde{\lambda}_{\dot{\alpha}}^i$,  $p_i^2=0$
carried by the on-shell particle. Then one can write the $n$-point
superamplitude as
\begin{equation}\label{superamplitude}
\mathcal{A}_n(\lambda,\tilde{\lambda},\eta)=\langle0|\prod_{i=1}^n\Omega_i|0\rangle,
\end{equation}
where the average $\langle0| \ldots |0\rangle$ is understood with
respect to some particular (for example component) formulation of
the $\mathcal{N}=4$ SYM theory. The "all ingoing" notation as well
as the color decomposition, the color ordering and the planar
limit\footnote{$g \rightarrow 0$ and $N_c \rightarrow \infty$ of
$SU(N_c)$ gauge group so that $\lambda=g^2N_c=$fixed.} are
implemented through out this paper for all objects containing
$|\Omega\rangle$ superstate.

We also use the standard $\mathcal{N}=4$ coordinate superspace,
which is convenient for the description of supermultiplets of fields
or operators and which is parameterized by the following
coordinates:
\begin{equation}
\mbox{$\mathcal{N}=4$ coordinate
superspace}=\{x^{\alpha\dot{\alpha}},~\theta^A_{\alpha},~\bar{\theta}_{A\dot{\alpha}}\},
\end{equation}
where $x_{\alpha\dot{\alpha}}$ are bosonic coordinates and
$\theta$'s, which are $SU(4)_R$ vectors and Lorentz $SL(2,C)$
spinors, are fermionic ones. The
generators of supersymmetry algebra relevant for our discussion are realized in this superspace
as
\begin{eqnarray}\label{SUSYGenCoord}
\mbox{4 translations }P_{\alpha\dot{\alpha}} &=& q_{\alpha\dot{\alpha}},\nonumber\\
\mbox{8 supercharges } Q_{\alpha}^A &=&
\frac{\partial}{\partial\theta^{\alpha}_A }
-\bar{\theta}^{\dot{\alpha }A}q_{\alpha\dot{\alpha}},\nonumber\\
\mbox{8 conjugated supercharges } \bar{Q}_{A\dot{\alpha}} &=&
-\frac{\partial}{\partial\bar{\theta}^{A\dot{\alpha} } } +
\theta^{\alpha}_{A}q_{\alpha\dot{\alpha}},
\end{eqnarray}
where one considers the Fourier transformed generators for
bosonic coordinates $x^{\alpha\dot{\alpha}}\rightarrow
q_{\alpha\dot{\alpha}}$.

The full $\mathcal{N}=4$ coordinate superspace is obviously
non-chiral in contrast to the on-shell momentum superspace. The
$\mathcal{N}=4$ supermultiplet of fields (containing $\phi^{AB}$
scalars, $\psi^A_{\alpha}, \bar{\psi}^A_{\dot{\alpha}}$ fermions and
$F^{\mu\nu}$-- the gauge field strength tensor, all in the adjoint
representation of $SU(N_c)$ gauge group) is realized in
$\mathcal{N}=4$ coordinate superspace as a constrained superfield $
W^{AB}(x,\theta,\bar{\theta})$ with the lowest component $
~W^{AB}(x,0,0)=\phi^{AB}(x)$. $W^{AB}$ in general is not a chiral
object and satisfies several constraints: a self-duality constraint
\begin{equation}
W^{AB}(x,\theta,\bar{\theta})
=\overline{W_{AB}}(x,\theta,\bar{\theta})=
\frac{1}{2}\epsilon^{ABCD}W_{CD}(x,\theta,\bar{\theta}),
\end{equation}
which implies $\phi^{AB}=\overline{\phi_{AB}}=\frac 12
\epsilon^{ABCD} \phi_{CD}$ and two additional
constraints\footnote{$[\ast,\star]$ denotes antisymmetrization in
indices, while $(\ast,\star)$ denotes symmetrization in indices.}
\begin{eqnarray}\label{PartialChirality} &&
D_C^{\alpha}W^{AB}(x,\theta,\bar{\theta}) = -\frac{2}{3}\delta^{[A}_CD_L^{\alpha}W^{B]L}(x,\theta,\bar{\theta}), \nonumber\\
&&\bar{D}^{\dot{\alpha}(C}W^{A)B}(x,\theta,\bar{\theta}) = 0,
\end{eqnarray}
where $D^A_{\alpha}$ is a standard coordinate superspace
derivative\footnote{which is
$D^A_{\alpha}=\partial/\partial\theta_{A}^{\alpha}
+i\bar{\theta}^{A\dot{\alpha}}\partial/\partial
x^{\alpha\dot{\alpha}}$.}. Note that in this formulation the full
$\mathcal{N}=4$ supermultiplet of fields is on-shell in the sense
the algebra (more precisely the last two anti commutators) of the
generators $Q^{A}_{\alpha},\bar{Q}_{B\dot{\alpha}}$ for the
supersymmetric transformation of the fields in this supermultiplet
\begin{equation}
\{Q^{A}_{\alpha},\bar{Q}_{B\dot{\alpha}}\}=2\delta^A_BP_{\alpha\dot{\alpha}},~
\{Q^{A}_{\alpha},Q^{B}_{\beta}\}=0,~\{\bar{Q}_{A\dot{\alpha}},\bar{Q}_{B\dot{\beta}}\}=0
\end{equation}
is closed only if the fields obey their equations of motion (in
addition the closure of the algebra requires the compensating gauge
transformation \cite{SuperCor1}).

Using (\ref{PartialChirality}) one can write \cite{SuperCor1}  any
$W^{AB}$ for a particular choice of $A,B$ in the form that  is independent of
 half of  $\theta$'s and
$\bar{\theta}$'s. This property is invariant under the
transformations of $\theta$'s and $\bar{\theta}$'s with respect to
some $SU(4)_R$ subgroup, while for all other values of $A$ and $B$
$W^{AB}$ contains the  full dependence on $\theta$'s and
$\bar{\theta}$'s. More accurately, one  breaks $SU(4)_R$ group into
two $SU(2)$ and $U(1)$
\begin{eqnarray}\label{projection1}
SU(4)_R &\rightarrow& SU(2)\times SU(2)'\times U(1),
\end{eqnarray}
so that the index $A$ of $R$-symmetry group $SU(4)_R$ splits into
\begin{eqnarray}\label{prrojection2}
A&\rightarrow&(+a|-a'),
\end{eqnarray}
where $+a$ and $-a'$ corresponds to two copies of $SU(2)$ and $\pm$
corresponds to the $U(1)$ charge (we do not write the  $U(1)$ factor
explicitly hereafter, and  use the notation $+a\equiv a$ and
$-a'\equiv\dot{a}$); and then takes particular ($ab$) projection of
$W^{AB}$ after which the constraints (\ref{PartialChirality}) become
\cite{SuperCor1}
\begin{eqnarray}\label{GrassmannianAnalit}
D_{\alpha}^{\dot{c}}W^{ab}(x,\theta,\bar{\theta}) &=& 0,\nonumber\\
\bar{D}_{c\dot{\alpha}}W^{ab}(x,\theta,\bar{\theta}) &=& 0.
\end{eqnarray}

The latter constraints are called
the Grassmannian analyticity conditions (see \cite{SuperCor1} and
references therein for details) resolving which one gets
\begin{equation}
W^{ab}(x,\theta,\bar{\theta}) =
W^{ab}(X,\theta^{c},\bar{\theta}_{\dot{c}}),
\end{equation}
where $X$ is a chiral coordinate
\begin{equation}
X^{\alpha\dot{\alpha}}=x^{\alpha\dot{\alpha}}
+i(\theta^{a}\bar{\theta}_{a})^{\alpha\dot{\alpha}}
-i(\theta^{\dot{a}}\bar{\theta}_{\dot{a}})^{\alpha\dot{\alpha}}.
\end{equation}

The full $\mathcal{N}=4$ supermultiplet can be truncated to the
so-called $\mathcal{N}=4$ vector chiral multiplet
$\mathcal{W}^{AB}$, which is closed under the chiral part of
supersymmetry generators $Q^{A}_{\alpha}$ and contains the self-dual
part of the $\mathcal{N}=4$ supermultiplet: $\phi^{AB}$ scalars,
$\psi_{\alpha}^A$ fermions and $F^{\alpha\beta}$ that is a
self-dual part of $F^{\mu\nu}$ defined as
$$
F^{\mu\nu} \rightarrow
F^{\alpha\dot{\alpha}\beta\dot{\beta}}=\epsilon^{\alpha\beta}\tilde{F}^{\dot{\alpha}\dot{\beta}}
+ \epsilon^{\dot{\alpha}\dot{\beta}}F^{\alpha\beta}.
$$
$\mathcal{W}^{AB}$ can be obtained by setting $\bar{\theta}=0$ in
$W^{AB}$. For the $\mathcal{W}^{ab}$ component (projection) of
$\mathcal{W}^{AB}$ one gets the following expansion in the Grassmannian
coordinates \cite{SuperCor1}:
\begin{equation}\label{Wchiral}
\mathcal{W}^{ab}(x,\theta^{c})=W^{ab}(x,\theta^{c},0)= \phi^{ab} -
\textup{i} \sqrt{2}(\theta^{[a}\psi^{b]}) +
\frac{\textup{i}}{\sqrt{2}} (\theta^{[a}\theta^{b]}F)+O(g).
\end{equation}
Here $O(g)$ corresponds to the terms proportional to commutators of
$SU(N_c)$ matrices in the fundamental representation and $(\ldots)$
stands for the contractions of $SL(2,C)$ indices. Note that in the
latter expression one can write $\phi^{ab}$ as
$\phi_{\dot{a}\dot{b}}$ due to the self-duality constraint. We also
want to point out that the component fields in $\mathcal{W}^{AB}$
are off-shell \cite{SuperCor1}, i.e. the component fields in
$\mathcal{W}^{AB}$ are arbitrary and the chiral part of the algebra
$\{Q^{A}_{\alpha},Q^{B}_{\beta}\}=0$ of  supersymmetric transformations of the component
fields in $\mathcal{W}^{AB}$ can be still closed without any
constraints on the component fields.

\subsection{The structure of the stress-tensor supermultiplet
in $\mathcal{N}=4$ coordinate superspace}

Let us now briefly discuss the structure of the stress-tensor
supermultiplet and the half-BPS multiplets in $\mathcal{N}=4$
superspace. The half-BPS multiplets by definition are annihilated by
half of supersymmetry generators of the theory. The simplest example
of the half-BPS operators in the $\mathcal{N}=4$ SYM theory is the
 lowest component operators in the supermultiplet which consists of $n$
identical scalars
\begin{equation}
\mathcal{O}^{(n)}_{AB}=Tr\left(\phi_{AB}^n\right).
\end{equation}

In the case of $n=2$ the operator $\mathcal{O}^{(2)}_{AB}$ belongs also to
the stress-tensor supermultiplet $T^{AB}$. This supermultiplet
contains the stress-tensor and all other conserved currents in the
$\mathcal{N}=4$ SYM theory. The stress-tensor supermultiplet is
realized as a superfield $T^{AB}$ in $\mathcal{N}=4$ coordinate
superspace\footnote{more accurately, one has to write
$T^{ABCD}=\mbox{Tr}\left(W^{AB}W^{CD} - \frac{1}{24}
\epsilon^{ABCD}\epsilon_{IKLM}W^{IK}W^{LM}\right)$.}
\begin{equation}
 T^{AB}=\mbox{Tr}\left(W^{AB}W^{AB}\right).
\end{equation}

The following notation $T^{(0)}_{AB}=Tr(\phi_{AB}^2)$ is used for
the lowest component of $T_{AB}$. The superfield $T^{AB}$ is not
chiral in contrast to $\langle\Omega_n|$ superstate. However, one
can restrict oneself to the chiral sector $\mathcal{T}^{AB}$ of
$T^{AB}$ which is realized for the particular projection
$\mathcal{T}^{ab}$ of $\mathcal{T}^{AB}$ as
\begin{equation}
\mathcal{T}^{ab}(x,\theta^c) =
Tr(\mathcal{W}^{ab}\mathcal{W}^{ab})(x,\theta^c),
~\mathcal{T}^{ab}(x,0)=Tr((\phi^{ab})^2).
\end{equation}
All the component fields in $\mathcal{T}^{ab}$, as was explained
earlier, are off-shell and $\mathcal{T}^{ab}$ is a pure chiral
object.

Using the supercharges as translation generators in superspace, one
can write $\mathcal{T}^{ab}$  as
\begin{equation}\label{TexpT0}
\mathcal{T}^{ab}(x,\theta^{c}) = \exp(\theta_{c}^{\alpha}Q^{c}_{\alpha})\mathcal{T}^{ab}(x,0).
\end{equation}
One would like to note that the self-duality constraint implies
$T^{(0)ab}=T^{(0)}_{\dot{a}\dot{b}}$.

Let us write explicitly the condition that operator
$T^{(0)}_{\dot{a}\dot{b}}$ as the member of the half-BPS multiplet is
annihilated by half of the SUSY generators. For
$T^{(0)}_{\dot{a}\dot{b}}$ being a projection of $T^{(0)}_{AB}$ one
splits supersymmetry generators as
\begin{eqnarray}
&& Q_{\alpha}^A\rightarrow (Q_{\alpha}^{a}|Q_{\alpha}^{\dot{a}});
\nonumber \\ && \bar{Q}_{A\dot{\alpha}}\rightarrow
(\bar{Q}_{a\dot{\alpha}}|\bar{Q}_{\dot{a}\dot{\alpha}}),
\end{eqnarray}
and skipping the Lorentz indices one has
\begin{eqnarray}\label{BPScondition}
&&[Q^{\dot{c}},T^{(0)}_{\dot{a}\dot{b}}]=0,~[\bar{Q}_{c},T^{(0)}_{\dot{a}\dot{b}}]=0.
\end{eqnarray}

\subsection{The MHV superamplitudes at the tree level}
Let us briefly discuss the amplitudes in on-shell momentum
superspace. The symmetry arguments and the component answer for the
tree-level  gluon MHV amplitudes  completely fix the MHV part of the
superamplitude (\ref{superamplitude}) at the tree level. One can use
it for the loop computations by means of the unitarity based method
and/or for the recursion relations to construct the non-MHV tree
amplitudes. The explicit form of the MHV tree amplitude is important
since the unitarity cut constructibility technique uses it as an
input for reconstruction of the loop amplitudes.

The superamplitude must be invariant under the transformations
generated by the full supersymmetry algebra in general, and in
particular under the (super)translations (\ref{SUSYGenLightCone})
$p_{\alpha\dot{\alpha}},~q_{\alpha}^A,~\bar{q}_{A\dot{\alpha}}$
which imply
\begin{equation}
p_{\alpha\dot{\alpha}}\mathcal{A}_n
=q_{\alpha}^A\mathcal{A}_n=\bar{q}_{A\dot{\alpha}}\mathcal{A}_n=0,
\end{equation}
where
$$
p_{\alpha\dot{\alpha}}=
\sum_{i=1}^n\lambda_{\alpha}^i\tilde{\lambda}_{\dot{\alpha}}^i,~
q_{\alpha}^{A}=\sum_{i=1}^n\lambda_{\alpha}^i\eta^A_i,~
\bar{q}_{A\dot{\alpha}}=\sum_{i=1}^n\tilde{\lambda}_{\dot{\alpha}}^i\frac{\partial}{\partial\eta^A_i}.
$$

From the above requirements one finds out that
\cite{DualConfInvForAmplitudesCorch}
\begin{equation}\label{superamplitude1}
\mathcal{A}_n(\lambda,\tilde{\lambda},\eta)=\delta^4(p_{\alpha\dot{\alpha}})
\delta^8(q_{\alpha}^A)\mathcal{P}_n(\lambda,\tilde{\lambda},\eta).
\end{equation}
Here $q_{\alpha}^A$ can be understood as a  superpartner of
momentum $p_{\alpha\dot{\alpha}}$. The fact that $\mathcal{A}_n
\sim \delta^4(\ldots) \delta^8(\ldots)$ is a reflection of
supermomentum conservation, i.e., translation invariance in momentum
superspace.  The Grassmannian
delta-function $\delta^{8}(q_{\alpha}^A)$ is defined by
\begin{equation}\label{delta8}
\delta^{8}(q_{\alpha}^A)=\sum_{i,j=1}^{n}\prod_{A,B=1}^{4}\langle
ij\rangle\eta^A_{i}\eta^{B}_{j}.
\end{equation}

The amplitude $\mathcal{A}_n$ is a polynomial in $\eta$ of the order of
$4n-8$ \cite{DualConfInvForAmplitudesCorch},
$\delta^8(q_{\alpha}^A)$ is a polynomial in $\eta$ of the order 8. Hence, if one
expands $\mathcal{P}_n$ in powers of $\eta$, one obtains
\begin{equation} \label{ExpansP}
\mathcal{P}_n = \mathcal{P}^{(0)}_n + \mathcal{P}^{(4)}_n + \ldots +
\mathcal{P}^{(4n-16)}_n,
\end{equation}
where $\mathcal{P}^{(4m)}_n$ is a homogenous $SU(4)_R$ invariant
polynomial of the order of $4m$ (it can be deduced from the requirement
$\bar{q}_{B\dot{\alpha}}\mathcal{A}_n=0$). If one assigns
the  helicity $\lambda=+1$ to $|\Omega_i\rangle$ and $\lambda=+1/2$ to
$\eta$ then the amplitude $\mathcal{A}_n$ has an overall helicity
$\lambda_{\Sigma}=n$, $\delta^8(q_{\alpha}^A)$  has helicity
$\lambda_{\Sigma}=4$ so that $\mathcal{P}^{(0)}_n$ has helicity
$\lambda_{\Sigma}=n-4$. The MHV superamplitude is
\begin{equation}\label{MHVsuperamplitude}
\mathcal{A}_n^{MHV}(\lambda,\tilde{\lambda},\eta)=\delta^4(p_{\alpha\dot{\alpha}})
\delta^8(q_{\alpha}^A)\mathcal{P}^{(0)}_n(\lambda,\tilde{\lambda}).
\end{equation}

The coefficients in the expansion of $\mathcal{A}_n^{MHV}$ with
respect to the Grassmannian variables $\eta^A_i$ are the component
amplitudes. For example,
$$
\mathcal{A}_n^{MHV}(\lambda,\tilde{\lambda},\eta) =
(\eta^1\eta^2\eta^3\eta^4)_1 (\eta^1\eta^2\eta^3\eta^4)_2\langle
g^-_1g^-_2g^+_3  \ldots g^+_n\rangle + \ldots,
$$
where $\langle g^-_1g^-_2g^+_3 \ldots g^+_n\rangle$ is the MHV
gluon component (Parke-Taylor) amplitude with all but two gluons
with positive helicities. To extract the component amplitude from
$\mathcal{A}_n^{MHV}$, it is convenient to define the projecting
operators
\begin{eqnarray}\label{projectors}
&&|g^+\rangle:~D^{+}_i=1,\nonumber\\&&|\Gamma^A\rangle:~D^{A}_i=\partial^{A}_i,\nonumber\\&&|\phi^{AB}\rangle:~D^{AB}_i=\frac{1}{2}\partial^{A}_i\partial^{B}_i,\nonumber\\
&&|\bar{\Gamma}^A\rangle:~\bar{D}^{D}_i=\frac{1}{3!}\varepsilon^{ABCD}\partial_{iA}\partial_{iB}\partial_{iC},\nonumber\\&&
|g^-\rangle:~D^{-}_i=\frac{1}{4!}\varepsilon^{ABCD}\partial_{iA}\partial_{iB}\partial_{iC}\partial_{iD},
\end{eqnarray}
which are the projectors on
$|g^+\rangle$,$|\Gamma^A\rangle$,$|\phi^{AB}\rangle$,$|\bar{\Gamma}^A\rangle$,$|g^-\rangle$
states, respectively. The label $i$  specifies the momentum of the
state. All the previous discussions are valid for both the tree and the loop
amplitudes.

One can determine the precise form of $\mathcal{P}^{(0)}_n$ for the
tree amplitudes using  the knowledge of the one particular component
amplitude (for example, the Parke-Taylor one)
\cite{DualConfInvForAmplitudesCorch}. Acting by the corresponding
projection operators (\ref{projectors}) on the MHV part of the
superamplitude
\begin{eqnarray}
&&\langle g^-_1g^-_2g^+_3\ldots
g^+_n\rangle^{tree}=D^{-}_1D^{-}_2D^{+}_3\ldots
D^{+}_n\mathcal{A}_n^{tree,~MHV}|_{\eta=0}=
\nonumber\\
&&(\partial^{1}\partial^{2}\partial^{3}\partial^{4})_1
(\partial^{1}\partial^{2}\partial^{3}\partial^{4})_2\delta^4(p_{\alpha\dot{\alpha}})
\delta^8(q_{\alpha}^A)\mathcal{P}_n^{(0)}|_{\eta=0}=
\delta^4(p_{\alpha\dot{\alpha}})\langle12\rangle^4\mathcal{P}_n^{(0)},
\end{eqnarray}
and comparing the result with the known component answer
\begin{equation}
\langle g^-_1g^-_2g^+_3 \ldots
g^+_n\rangle^{tree}=\delta^4(p_{\alpha\dot{\alpha}})\frac{\langle12\rangle^4}{\langle12\rangle\ldots
\langle n1\rangle},
\end{equation}
one gets
\begin{equation} \label{AmpMHVzero}
\mathcal{P}_n^{(0)} = \frac{1}{\langle12\rangle \ldots \langle
n1\rangle}.
\end{equation}
So for the MHV part of the superamplitude
\begin{equation}
\mathcal{A}_n^{tree,~MHV}(\lambda,\tilde{\lambda},\eta)=\delta^4(p_{\alpha\dot{\alpha}})
\delta^8(q_{\alpha}^A)\frac{1}{\langle12\rangle \ldots \langle
n1\rangle}.
\end{equation}

For the computations using the unitarity based techniques for the
MHV sub-sector and two-particle (iterated) cuts one needs only the
tree MHV superamplitudes $\mathcal{A}_n^{MHV,~tree}$. However, the
tree MHV amplitudes are also important for other calculations
involving the NMHV amplitudes \cite{SuperSums}. The fact that for
the MHV sector/two-particle (iterated) cuts only the MHV amplitudes
contribute can be seen from the na\"ive counting of  $\eta$'s
$$
\mathcal{A}_4^{1-loop,MHV}\sim\eta^4\eta^4,~dSLIPS_2\sim\partial^8/\partial\eta^8,
~~\mathcal{A}_4^{tree,MHV}\sim\eta^4\eta^4.
$$
Here the super-Lorentz invariant phase space $dSLIPS_n$ is
defined as
\begin{equation}
dSLIPS_n^{l_1, \ldots
,l_n}=\delta^4\left(\sum_{i=1}^nl_i+\sum_{k=1}^mp_{k}\right)
\prod_{i=1}^nd^4\eta_i~\delta^{+}(l_i^2)\frac{d^Dl_i}{(2\pi)^{D-1}},
\end{equation}
where $l_i, i=1,\ldots,n$ are the momenta crossed by  the cut with the
associated Grassmannian variables $\eta^A_i$ and $p_k, k=1, \ldots, m$
are the external momenta.

For example, for the one-loop two-particle cut in the $s_{ij}=(p_i+p_j)^2$
channel one has
\begin{equation}
disc_{s_{ij}}[\mathcal{A}_4^{1-loop,MHV}] = \int
dSLIPS_2^{l_1l_2}\mathcal{A}_4^{tree,MHV}\mathcal{A}_4^{tree,MHV}.
\end{equation}

The same is true for the iterated cuts with tree amplitudes at
higher order of PT. The latter fact resembles the MHV vertex
expansion method \cite{rec0}, where only the MHV amplitudes as
effective vertices in the diagrams are used. In the on-shell
momentum superspace for the two-particle cuts of the MHV part of the
amplitude the following formula  for summation over the states in
the cut (Grasmannian integration in $dSLIPS$) can be applied
\begin{eqnarray}\label{supersumm}
&& \int d^4\eta_{l_1}d^4\eta_{l_2} \delta^{8} \left(
\lambda_{\alpha}^{l_1}\eta_{l_1}^A+\lambda_{\alpha}^{l_2}\eta_{l_2}^A+Q_{\alpha}^A
\right)
\delta^{8} \left( \lambda_{\alpha}^{l_1}\eta_{l_1}^A+\lambda_{\alpha}^{l_2}\eta_{l_2}^A-P_{\alpha}^A \right) \nonumber\\
&& \makebox[7em]{} = \langle
l_1l_2\rangle^4\delta^8\left(P_{\alpha}^A+Q_{\alpha}^A\right).
\end{eqnarray}

Similar results are also obtained for the amplitudes in the theories
with lower supersymmetry ($1 \leq \mathcal{N} <4$) \cite{LessSUSY}.
For example, one can write an analog of (\ref{supersumm}) in theories
with $1 \leq \mathcal{N} <4$ supersymmetry \cite{LessSUSY}
\begin{eqnarray}\label{supersummGeneralTHeory}
&&\int d^\mathcal{N} \eta_{l_1} d^\mathcal{N}\eta_{l_2}
\delta^{2\mathcal{N}}
\left(\lambda_{\alpha}^{l_1}\eta_{l_1}^A+\lambda_{\alpha}^{l_2}\eta_{l_2}^A+Q_{\alpha}^A\right)
\delta^{2\mathcal{N}} \left(\lambda_{\alpha}^{l_1}\eta_{l_1}^A+\lambda_{\alpha}^{l_2}\eta_{l_2}^A-P_{\alpha}^A\right)\nonumber\\
&&=\langle l_1l_2\rangle^{\mathcal{N}}
\delta^{2\mathcal{N}}\left(P_{\alpha}^A+Q_{\alpha}^A\right),
\end{eqnarray}
where $A$ is the R-symmetry group index (for $SU(2)_R$ and $SU(4)_R$ in case of $\mathcal{N}=2$
and $\mathcal{N}=4$ SUSY, respectively). The
definition of the Grassmannian delta-function
$\delta^{2\mathcal{N}}(\ldots)$  for $\mathcal{N}=4$ is given by (\ref{delta8}), and for
$\mathcal{N}=2$ the definition is presented further in the text \cite{LessSUSY}.

It is possible to write an on-shell momentum superspace
generalization for form factors of chiral truncation for the
stress-tensor supermultiplet $T^{AB}$. Moreover, using this
formulation the supersums in unitarity based loop computations for
super form factors in the MHV sector can be performed in a similar
way as for the amplitudes.

\subsection{Superstate -- form factor}
To construct "superstate -- form factor" as a generalization of
component form factors with the $T^{(0)}_{AB}$ operator, one begins
with the observation that for the form factor of the operator
$T^{(0)}_{AB}$ with external state made of gluons with $+$
polarization and 2 scalars $\overline{\phi_{AB}}$ the answer is
known at the tree level \cite{FormFMHV}. For some particular choice
of $A, B$ (let us take $\dot{a}, \dot{b}$) one gets
\begin{equation}\label{componentTravaglini}
\langle g^+_1 \ldots \phi_{i~ab} \ldots \phi_{j~ab} \ldots
g^+_n|T^{(0)}_{\dot{a}\dot{b}}(q)|0\rangle=
\delta^4(\sum_{l=1}^n\lambda_{\alpha}^l\tilde{\lambda}_{\dot{\alpha}}^l
- q_{\alpha\dot{\alpha}}) \frac{\langle
ij\rangle^2}{\langle12\rangle \ldots \langle n1\rangle},
\end{equation}
where $q$ is arbitrary momentum carried by the
$T^{(0)}_{\dot{a}\dot{b}}$ operator.

One considers a generalization of this result to the object of the
form
\begin{equation}\label{Fsuper-component0}
F_n(\{\lambda,\tilde{\lambda},\eta\},q)=\langle\Omega_n|T^{(0)}_{\dot{a}\dot{b}}(q)|0\rangle,
\end{equation}
where we keep the  operator in component form and promote the  external
state to its superversion (\ref{superstate})
\begin{equation}
\langle\Omega_n|=\langle0|\prod_{i=1}^n\Omega_i.
\end{equation}

Using the fact that the supercharges annihilate the vacuum and the commutation relations
 (\ref{BPScondition}) one gets
\begin{equation}
\langle\Omega_n|Q_{\alpha}^{\dot{c}}T^{(0)}_{\dot{a}\dot{b}}|0\rangle
=
\langle\Omega_n|\bar{Q}_{c\dot{\alpha}}T^{(0)}_{\dot{a}\dot{b}}|0\rangle=0.
\end{equation}
In combination with  momentum
conservation this leads to the following properties of the superstate form factor $F_n$ \footnote{We thank A. Zhiboedov
for discussion of this point.}:
\begin{equation}\label{symmetryConditionsForF}
q_{\alpha}^{\dot{a}}F_n=\bar{q}_{\dot{\alpha}a}F_n=p_{\alpha\dot{\alpha}}F_n=0,
\end{equation}
where
\begin{equation}
q_{\alpha}^{\dot{a}}=\sum_{i=1}^n\lambda_{\alpha}^i\eta^{\dot{a}}_i
,\
\bar{q}_{a\dot{\alpha}}=\sum_{i=1}^n\tilde{\lambda}_{\dot{\alpha}}^i\frac{\partial}{\partial\eta^a_i},
\  p_{\alpha\dot{\alpha}}=
\sum_{i=1}^n\lambda_{\alpha}^i\tilde{\lambda}_{\dot{\alpha}}^i-q_{\alpha\dot{\alpha}}.
\end{equation}

Then one writes, in full analogy with superamplitude
(\ref{superamplitude1}), the expression for the super form factor
\begin{equation}
F_n(\{\lambda,\tilde{\lambda},\eta\},q)=
\delta^4(\sum_{i=1}^n\lambda_{\alpha}^i\tilde{\lambda}_{\dot{\alpha}}^i-q_{\alpha\dot{\alpha}})
\delta^4_{GR}\left(q^{\dot{a}}_{\alpha}\right)
\left(\mathcal{X}_n^{(0)}+\mathcal{X}_n^{(4)}+ \ldots
+\mathcal{X}_n^{(4n-8)}\right)
\end{equation}
where $\mathcal{X}^{(4m)}_n$ are the homogenous $SU(4)_R$ invariant
polynomials of the order of $4m$. The Grassmannian delta-function
$\delta^4_{GR}$ is defined as (one uses the $GR$ subscript to
distinguish it from the ordinary bosonic delta-function)
\begin{equation}
\delta^4_{GR}\left(\sum_{i=1}^n\lambda_{\alpha}^i\eta^{\dot{a}}_i\right)
=\sum_{i,j=1}^{n} \prod_{\dot{a},\dot{b}=1,2} \langle
ij\rangle\eta^{\dot{a}}_i\eta^{\dot{b}}_j.
\end{equation}

Here in contrast to the expression (\ref{ExpansP}) for the
amplitudes where the expansion was up to the polynomial
$\mathcal{P}_n^{(4n-16)}$ one has only the expansion up to
$\mathcal{X}_n^{(4n-8)}$ associated with the difference in the
number of super charges which annihilate $F_n$.

If one, as in the case of superamplitude,  assigns helicity  $\lambda=+1$
to $|\Omega_i\rangle$ and $\lambda=+1/2$ to $\eta$, one
sees that $F_n$ has an overall helicity $\lambda_{\Sigma}=n$,
$\delta^4_{GR}$ has $\lambda_{\Sigma}=2$ so that $\mathcal{X}^{(0)}_n$
has $\lambda_{\Sigma}=n-2$ which is understood as an analog of the
MHV part of  superamplitude (\ref{superamplitude})
\begin{equation}
F_n^{MHV}(\{\lambda,\tilde{\lambda},\eta\},q)
=\delta^4(\sum_{i=1}^n\lambda_{\alpha}^i\tilde{\lambda}_{\dot{\alpha}}^i-q_{\alpha\dot{\alpha}})
\delta^4_{GR}\left(q^{\dot{a}}_{\alpha}\right)
\mathcal{X}^{(0)}_n(\lambda,\tilde{\lambda}).
\end{equation}

The non-MHV contributions ($\mbox{N}^k\mbox{MHV}$) to
$F_n(\{\lambda,\tilde{\lambda},\eta\},q)$ are not considered in this
paper. To obtain the component answer from the latter expression, the
projection operators (\ref{projectors}) are considered in full
analogy with the superamplitude. The overall order of
projectors in this case is 4 instead 8 for the superamplitude. At the
tree level, comparing this result with the component answer
(\ref{componentTravaglini}) one obtains
\begin{equation} \label{FFMHVzero}
\mathcal{X}^{(0)}_n = \frac{1}{\langle12\rangle \ldots \langle
n1\rangle}.
\end{equation}

Being manifestly not $SU(4)_R$ covariant $\delta^4_{GR}$ is not
suitable for performing the supersummations since it depends only on  half
of $\eta$ variables. In principle, one can formally write $\delta^4_{GR}$
as
\begin{equation}
\delta^4_{GR}(q^{\dot{a}}_{\alpha})=\frac{(\partial^{a}\partial^{b})^{'}
(\partial^{a}\partial^{b})^{''}}{\langle\lambda^{'}\lambda^{''}\rangle^2}
\delta^8(q^{A}_{\alpha}+\lambda_{\alpha}^{'}\eta^{'A}+\lambda_{\alpha}^{''}\eta^{''A})|_{\eta^{'}=\eta^{''}=0}.
\end{equation}
Defining a projection operator
\begin{equation}
\Pi^{\dot{a}\dot{b}}=\frac{(\partial^{a}\partial^{b})^{'}
(\partial^{a}\partial^{b})^{''}}{\langle\lambda^{'}\lambda^{''}\rangle^2}|_{\eta^{'}=\eta^{''}=0},
\end{equation}
one can then write the MHV part of the form factor at the tree level as
\begin{equation}\label{Fsuper-component}
F_n^{MHV,tree}(\{\lambda,\tilde{\lambda},\eta\},q)
=\delta^4(\sum_{i=1}^n\lambda_{\alpha}^i\tilde{\lambda}_{\dot{\alpha}}^i-q_{\alpha\dot{\alpha}})
\frac{\Pi^{\dot{a}\dot{b}}\delta^8\left(q^{A}_{\alpha}\right)}{\langle
12\rangle \ldots \langle n1\rangle},
\end{equation}
where now
$$
q^{A}_{\alpha}=\sum_{i=1}^n\lambda_{\alpha}^i\eta^{A}_i+
\lambda_{\alpha}^{'}\eta^{'A}+\lambda_{\alpha}^{''}\eta^{''A}.
$$

For the form factor written in such a form the two-particle supersums for the MHV
sector are performed in the same fashion as for the amplitudes since the auxiliary
variables $\eta^{'},\eta^{''}$ are always harmless and the projection operator $\Pi^{\dot{a}\dot{b}}$ stays
outside the Grassmannian integral in $dSLIPS$.

Acting on
(\ref{Fsuper-component}) with the projection operators
(\ref{projectors}) to external state one obtains the component form factors. For example, the form factor
of $T^{(0)}_{\dot{a}\dot{b}}$ with external state consisting of the
gluons with $+$ polarization, 1 scalar and 2 fermions is given by
$$
\langle g^+_1 \ldots \phi_{i~ab} \ldots \psi_{j~a}^{} \ldots
\psi_{k~b}^{} \ldots g^+_n|T^{(0)}_{\dot{a}\dot{b}}(q)|0\rangle=
\delta^4(\sum_{l=1}^n\lambda_{\alpha}^l\tilde{\lambda}_{\dot{\alpha}}^l
- q_{\alpha\dot{\alpha}}) \frac{\langle ij\rangle\langle
ik\rangle}{\langle12\rangle \ldots \langle n1\rangle}.
$$

For a small number of particles in external state ($n=3$, for example)
one easily verifies this result by direct diagram computation.
Formally, to obtain this answer for different values of $SU(4)_R$
indices of external state particles one has to consider different
projections; however, this expression is independent of the
particular values of the $SU(4)_R$ index.

In principle, expression (\ref{Fsuper-component}) can be used in the
unitarity based computations as a building block together with
$\mathcal{A}_n^{MHV,tree}$. However, it is desirable to obtain the
formulation of super form factors where the operator is also
promoted to its supersymmetric version. The next subsection is
devoted to such generalization.

\subsection{Superstate -- super form factor}
It is natural to consider the operator $T^{(0)}_{\dot{a}\dot{b}}$ as
the first term in series expansion in $\theta$'s of the
stress-tensor supermultiplet $T^{AB}$ for a particular choice of $A$
and $B$. However, as was mentioned before, $T^{AB}$ is not a chiral
superfield in contrast to the superstate $\langle\Omega_n|$ and it
is not convenient to consider the objects of different chirality
types. Also, the component fields in $T^{AB}$ are constrained, while
for insertion of an operator a pure off-shell object is preferred.
So one restricts oneself to the superfield $\mathcal{T}^{ab}$
instead of $T^{AB}$; $T^{(0)}_{\dot{a}\dot{b}}$ is also the first
term of the expansion in $\theta$'s of $\mathcal{T}^{ab}$ which
contains in particular the chiral part of projection
(\ref{projection1}), (\ref{prrojection2}) of the supermultiplet
$T^{AB}$ and is closed under the chiral $\mathcal{N}=4$
supersymmetric transformations. Here $\mathcal{T}^{ab}$ is used as
the operator insertion. The generalization of
(\ref{Fsuper-component0}) would be the replacement of
$T^{(0)}_{\dot{a}\dot{b}}$ in (\ref{Fsuper-component0}) by
$\mathcal{T}^{ab}$. This is because all the fields in
$\mathcal{T}^{ab}$ are still unconstrained (off-shell).

Thus, we consider a more general object different from
(\ref{Fsuper-component0}) of the following form:
$$
\mathcal{F}_n(\{\lambda,\tilde{\lambda},\eta\},q,\theta^{a})=
\langle\Omega_n|\mathcal{T}^{ab}(q,\theta^{a})|0\rangle.
$$

Remind that  $\mathcal{F}_n$ depends only on $\theta^{a}, a=1,2$ and
not on $\theta^{\dot{a}}, \dot{a}=1,2$. Using (\ref{TexpT0}) we can
rewrite it explicitly as
\begin{equation}\label{componentFormFactor}
\mathcal{F}_n(\{\lambda,\tilde{\lambda},\eta\},q,\theta^{a}) =
\exp(\theta_{a}^{\alpha}q^{a}_{\alpha})\langle\Omega_n|T^{(0)}_{\dot{a}\dot{b}}(q)|0\rangle
=
\exp(\theta_{a}^{\alpha}q^{a}_{\alpha})F_n(\{\lambda,\tilde{\lambda},\eta\},q),
\end{equation}
where $q^{a}_{\alpha}=\sum_{i=1}^n\lambda_{\alpha}^i\eta^a$. Hence,
following the  discussion in the previous section one gets
\begin{equation}
\mathcal{F}_n(\{\lambda,\tilde{\lambda},\eta\},q,\theta^{a}) =
e^{\theta_{a}^{\alpha}q^{a}_{\alpha}} \delta^4(\sum_{i=1}^n
\lambda_{\alpha}^i \tilde{\lambda}_{\dot{\alpha}}^i -
q_{\alpha\dot{\alpha}})
\delta^4_{GR}\left(q^{\dot{a}}_{\alpha}\right)
\left(\mathcal{X}_n^{(0)} + \mathcal{X}_n^{(4)} + \ldots +
\mathcal{X}_n^{(4n-8)}\right).
\end{equation}

We see that the generalization of the previous results to the
"superstate-- super form factor" is straightforward. Let us make a
comment about the helicity structure of $\mathcal{F}_n$. If we
assign helicity $\lambda=-1/2$ to $\theta_{\alpha}^{a}$  then the
exponential factor carries the total helicity $\lambda_{\Sigma}=0$.
Using the arguments from the previous subsection we conclude that
$\mathcal{X}_n^{(0)}$ still has the total helicity $\lambda_{\Sigma}
= n-2$ and it is understood as an analog of the MHV part for the
superamplitude. At the tree level, comparing the obtained expression
with the component answer (\ref{componentFormFactor}) and
identifying the first term in $\theta$ expansion with
(\ref{componentFormFactor}) one obtains
$\mathcal{X}_n^{(0)}=1/\langle12\rangle \ldots \langle n1\rangle$ so
that for the MHV sector at the tree level one gets
\begin{equation}
\mathcal{F}_n^{MHV,tree}(\{\lambda,\tilde{\lambda},\eta\},q,\theta^{a})=
\delta^4 (\sum_{i=1}^n \lambda_{\alpha}^i
\tilde{\lambda}_{\dot{\alpha}}^i - q_{\alpha\dot{\alpha}})
\exp(\theta_{a}^{\alpha}q^{a}_{\alpha}) \frac{\delta^4_{GR} \left(
q^{\dot{a}}_{\alpha} \right)}{\langle12\rangle \ldots \langle
n1\rangle}.
\end{equation}
Note that similar structures, namely,  the exponent
$\exp(\theta_{a}^{\alpha}q^{a}_{\alpha})$ in the context
of form factors in $\mathcal{N}=4$ SYM first appeared in
\cite{Perturbiner} where the non-gauge invariant form factors ("off
shell currents") of the form $\langle\Omega_n|W^{AB}|0\rangle$ were
studied.

Expanding $\exp(\theta_{a}^{\alpha}q^{a}_{\alpha})$ in powers of
$\theta$ one gets for the $m$'s ($m \leq 4$) coefficient of
expansion of the form factors in the form
$\langle\Omega_n|T^{ab}_{(m)}|0\rangle$, where $T^{ab}_{(m)}$ are
the operators which belong to the chiral part of the supermultiplet
$T^{AB}$. For example, the operators that are the Lorentz scalars have
the following form:
\begin{equation}\label{operatorsftomF}
T^{ab}_{(0)} = Tr(\phi_{\dot{a}\dot{b}}^2),~ T^{ab}_{(2)} =
Tr(\psi^a_{\alpha}\psi^{a\alpha}),~ T^{ab}_{(4)} =
Tr(F^{\alpha\beta}F_{\alpha\beta})+O(g).
\end{equation}

One can also assign a projection operator of the form
$\partial_{\theta}^m|_{\theta=0}$ to each of the form factors
$\langle\Omega_n|T^{ab}_{(m)}|0\rangle$. If one wants to obtain
operators with different values of indices from the chiral part of
$T^{AB}$, one has to consider different projections.

Let us now discuss the properties of $\mathcal{F}_n^{tree,MHV}$
under supersymmetric transformations in more detail. The part of
supersymmetry generators (\ref{SUSYGenCoord}) which acts on
$\mathcal{T}^{ab}$ can be written as
\begin{eqnarray}\label{SUSYGenCoordChiral}
\mbox{4 translations } P_{\alpha\dot{\alpha}}&=&q_{\alpha\dot{\alpha}},\nonumber\\
\mbox{8 supercharges } Q_{\alpha}^A&=&\frac{\partial}{\partial\theta^{\alpha}_A },\nonumber\\
\mbox{8 conjugated supercharges }
\bar{Q}_{A\dot{\alpha}}&=&\theta^{\alpha}_{A}q_{\alpha\dot{\alpha}},
\end{eqnarray}
while the supersymmetry generators acting on
$\mathcal{F}_n^{tree,MHV}$ are
\begin{eqnarray}\label{SUSYGenF}
\mbox{4 translations } P_{\alpha\dot{\alpha}}&=&
-\sum_{i=1}^n\lambda_{\alpha}^i\tilde{\lambda}_{\dot{\alpha}}^i+q_{\alpha\dot{\alpha}},\nonumber\\
\mbox{4 supercharges }
Q_{\alpha}^a&=&-\sum_{i=1}^n\lambda_{\alpha}^i\eta^{a}_i
+\frac{\partial}{\partial\theta^{\alpha}_a },\nonumber\\
\mbox{4 supercharges }
Q_{\alpha}^{\dot{a}}&=&-\sum_{i=1}^n\lambda_{\alpha}^i\eta^{\dot{a}}_i
+ \frac{\partial}{\partial\theta^{\alpha}_{\dot{a}} },\nonumber\\
\mbox{4 conjugated supercharges }
\bar{Q}_{a\dot{\alpha}}&=&-\sum_{i=1}^n\tilde{\lambda}_{\dot{\alpha}}^i
\frac{\partial}{\partial\eta^{a}_i}+\theta^{\alpha}_{a}q_{\alpha\dot{\alpha}},\nonumber\\
\mbox{4 conjugated supercharges }
\bar{Q}_{\dot{a}\dot{\alpha}}&=&-\sum_{i=1}^n\tilde{\lambda}_{\dot{\alpha}}^i
\frac{\partial}{\partial\eta^{\dot{a}}_i}+\theta^{\alpha}_{\dot{a}}q_{\alpha\dot{\alpha}}.
\end{eqnarray}

The action of the supersymmetric generators $Q_{\alpha}^a$ and
$Q_{\alpha}^{\dot{a}}$ on $\mathcal{F}_n^{tree,MHV}$ is given by
\begin{eqnarray}\label{eq1}
&&Q_{\alpha}^a\mathcal{F}_n^{tree,MHV} =
\delta^4(\sum_{i=1}^n\lambda_{\alpha}^i\tilde{\lambda}_{\dot{\alpha}}^i-q_{\alpha\dot{\alpha}})
\left(Q_{\alpha}^ae^{\theta_{c}^{\beta} q^{c}_{\beta}}\right)
\frac{\delta^4_{GR}(\sum_{i=1}^n\lambda_{\alpha}^i\eta^{\dot{a}}_i)}{\langle12\rangle
\ldots \langle
n1\rangle}= \nonumber\\
&& =
\delta^4(\sum_{i=1}^n\lambda_{\alpha}^i\tilde{\lambda}_{\dot{\alpha}}^i-q_{\alpha\dot{\alpha}})
\left(-\sum_{i=1}^n\lambda_{\alpha}^i\eta^a_i +
\sum_{i=1}^n\lambda_{\alpha}^i\eta^a_i\right)
e^{\theta_{c}^{\beta}q^{c}_{\beta}}
\frac{\delta^4_{GR}(\sum_{i=1}^n\lambda_{\alpha}^i\eta^{\dot{a}}_i)}{\langle12\rangle
\ldots \langle n1\rangle}=0,
\end{eqnarray}
\begin{eqnarray}\label{eq2}
Q_{\alpha}^{\dot{a}}\mathcal{F}_n^{tree,MHV} =
\delta^4(\sum_{i=1}^n\lambda_{\alpha}^i\tilde{\lambda}_{\dot{\alpha}}^i-q_{\alpha\dot{\alpha}})
\left(-\sum_{i=1}^n\lambda_{\alpha}^i\eta^{\dot{a}}_i\right)
\delta^4_{GR}(\sum_{i=1}^n\lambda_{\alpha}^i\eta^{\dot{a}}_i)
\frac{e^{\theta_{c}^{\beta}q^{c}_{\beta}}}{\langle12\rangle \ldots
\langle n1\rangle}=0.
\end{eqnarray}

From (\ref{eq1}) and (\ref{eq2}) one observes that
$\mathcal{F}_n^{tree,MHV}$ is invariant under the full chiral part
of supersymmetry generators $Q_{\alpha}^{A}$
$$Q_{\alpha}^{A}\mathcal{F}_n^{tree,MHV}=0.$$ This might be little
surprising since one expects that the half-BPS objects should be
annihilated only by half of chiral and antichiral supercharges.
Heuristically, this can be explained as reflection of the fact that
$\mathcal{F}_n$ is not only a half-BPS object but also contains an
operator from chiral projection of $T^{AB}$ which should be closed
under the chiral part of supersymmetry.

The action of antichiral supersymmetric generators is
\begin{eqnarray}
&&\bar{Q}_{a\dot{\alpha}}\mathcal{F}_n^{tree,MHV}=
\delta^4(\sum_{i=1}^n\lambda_{\alpha}^i\tilde{\lambda}_{\dot{\alpha}}^i-q_{\alpha\dot{\alpha}})\left(\bar{Q}_{a\dot{\alpha}}e^{\theta_{c}^{\beta}q^{c}_{\beta}}\right)
\frac{\delta^4_{GR}(\sum_{i=1}^n\lambda_{\alpha}^i\eta^{\dot{a}}_i)}{\langle12\rangle
\ldots \langle
n1\rangle}=\nonumber\\&&\delta^4(\sum_{i=1}^n\lambda_{\alpha}^i\tilde{\lambda}_{\dot{\alpha}}^i-q_{\alpha\dot{\alpha}})\
\left(-\sum_{i=1}^n\tilde{\lambda}_{\dot{\alpha}}^i\theta^{\beta}_{a}\lambda_{\beta}^i
+\theta^{\beta}_aq_{\beta\dot{\alpha}}\right)e^{\theta_{c}^{\beta}q^{c}_{\beta}}
\frac{\delta^4_{GR}(\sum_{i=1}^n\lambda_{\alpha}^i\eta^{\dot{a}}_i)}{\langle12\rangle
\ldots \langle n1\rangle}=\nonumber\\
&&\delta^4(\sum_{i=1}^n\lambda_{\alpha}^i\tilde{\lambda}_{\dot{\alpha}}^i-q_{\alpha\dot{\alpha}})\sum_{i=1}^n\theta^{\beta}_a
\left(-\tilde{\lambda}_{\dot{\alpha}}^i\lambda_{\beta}^i+
\tilde{\lambda}_{\dot{\alpha}}^i\lambda_{\beta}^i\right)e^{\theta_{c}^{\beta}q^{c}_{\beta}}\frac{\delta^4_{GR}(\sum_{i=1}^n\lambda_{\alpha}^i\eta^{\dot{a}}_i)}{\langle12\rangle
\ldots \langle n1\rangle}=0,
\end{eqnarray}
while for $\bar{Q}_{\dot{a}\dot{\alpha}}$
\begin{eqnarray}\label{anomaly}
&&\bar{Q}_{\dot{a}\dot{\alpha}}\mathcal{F}_n^{tree,MHV}=
\delta^4(\sum_{i=1}^n\lambda_{\alpha}^i\tilde{\lambda}_{\dot{\alpha}}^i-q_{\alpha\dot{\alpha}})\left(\bar{Q}_{\dot{a}\dot{\alpha}}\delta^4_{GR}(\sum_{i=1}^n\lambda_{\alpha}^i\eta^{\dot{a}}_i)\right)
\frac{e^{\theta_{a}^{\beta}q^{a}_{\beta}}}{\langle12\rangle \ldots
\langle n1\rangle}= \nonumber\\&& \makebox[-0.2em]{}
\delta^4(\sum_{i=1}^n\lambda_{\alpha}^i\tilde{\lambda}_{\dot{\alpha}}^i-q_{\alpha\dot{\alpha}})
\left(-q_{\alpha\dot{\alpha}}+\delta^4_{GR}(\sum_{i=1}^n
\lambda_{\alpha}^i\eta^{\dot{a}}_i)\theta_{\dot{a}}^{\beta}q_{\beta\dot{\alpha}}\right)
\frac{e^{\theta_{a}^{\beta}q^{a}_{\beta}}}{\langle12\rangle \ldots
\langle n1\rangle}\neq0.
\end{eqnarray}
So $\mathcal{F}_n^{tree,MHV}$ is invariant under half of the
antichiral supersymmetry generators $\bar{Q}_{a\dot{\alpha}}$, which
reflects the fact that $\mathcal{F}_n^{tree,MHV}$ is a half-BPS
object.

$\mathcal{F}_n^{MHV,tree}$ being a pure chiral object can be
rewritten in the form which resembles the superamplitude given by
(\ref{superamplitude1}). For this one should define the
transformation in the same spirit as \cite{TwistorTransform}, but in
this case the logic is inverted, and one replaces  a "physical"
coordinate $\theta$ with the set of auxiliary variables (moduli)
$\{\lambda,\eta\}$
\begin{equation}
\hat{T}[\ldots] = \int d^4\theta^{a}_{\alpha}
\exp(\theta_{a}^{\alpha} \sum_{i=1}^n
\lambda_{\alpha}^i\eta^{a}_i)[\ldots].
\end{equation}

For $n=2$ (for $n=1$ the Grassmannian delta function vanishes) one has
\begin{equation}
\hat{T}[1]=\delta^4_{GR}(\sum_{i=1}^2\lambda_{\alpha}^i\eta^{a}_i),
\end{equation}
and applying the transformation $\hat{T}$ to
$\mathcal{F}_n^{tree,MHV}$
\begin{eqnarray}\label{T[superFormfactor]}
&&Z^{tree,MHV}_n (\{\lambda,\tilde{\lambda},\eta\}, q,
\{\lambda{'},\lambda{''},\eta^{'a},\eta^{''a}\}) =
\hat{T}[\mathcal{F}_n^{tree,MHV}] = \nonumber\\
&&\delta^4(\sum_{i=1}^n\lambda_{\alpha}^i\tilde{\lambda}_{\dot{\alpha}}^i-q_{\alpha\dot{\alpha}})
\frac{\delta^4_{GR}(q^{a}_{\alpha} + \lambda^{'}_{\alpha}\eta^{'a} +
\lambda^{''}_{\alpha}\eta^{''a}) \delta^4_{GR}
\left(q^{\dot{a}}_{\alpha}\right)}{\langle12\rangle \ldots \langle
n1\rangle},
\end{eqnarray}
where $\{\lambda{'},\lambda{''},\eta^{'a},\eta^{''a}\}$ are the
auxiliary variables which replace the $\theta$ coordinate. This
expression is still not exactly $SU(4)_R$ covariant. However, one
can still perform the supersummation using
(\ref{supersummGeneralTHeory}). One can decompose the  Grassmannian
delta-function $\delta^8$ in the tree MHV amplitude into the product
of two $\delta^4_{GR}$ and the integration measure $d^4\eta^A
\rightarrow d^2\eta^ad^2\eta^{\dot{a}}$  and then use the relation
which consists of two copies of (\ref{supersummGeneralTHeory}),
where one has to put $\mathcal{N}=2$ (which corresponds to braking
of $SU(4)_R$ down to two $SU(2)$s in our case)
\begin{eqnarray}\label{FFsupersumm}
&&\int d^2\eta_{l_1}^{a}d^2\eta_{l_2}^{a}
d^2\eta_{l_1}^{\dot{a}}d^2\eta_{l_2}^{\dot{a}} \delta^{4}_{GR}
\left(\lambda_{\alpha}^{l_1}\eta_{l_1}^a
+\lambda_{\alpha}^{l_2}\eta_{l_2}^a +Q_{\alpha}^a\right)
\delta^{4}_{GR} \left( \lambda_{\alpha}^{l_1} \eta_{l_1}^{\dot{a}} +
\lambda_{\alpha}^{l_2}\eta_{l_2}^{\dot{a}} +
\tilde{Q}_{\alpha}^{\dot{a}}\right)\nonumber\\
&& \makebox[4em]{} \times
\delta^{4}_{GR}\left(\lambda_{\alpha}^{l_1}\eta_{l_1}^a +
\lambda_{\alpha}^{l_2}\eta_{l_2}^a-P_{\alpha}^a\right)
\delta^{4}_{GR} \left(\lambda_{\alpha}^{l_1}\eta_{l_1}^{\dot{a}} +
\lambda_{\alpha}^{l_2} \eta_{l_2}^{\dot{a}} -
P_{\alpha}^{\dot{a}}\right) \nonumber\\ && \makebox[6.35em]{} =
\langle l_1l_2\rangle^{4}\delta^{4}_{GR}\left(P_{\alpha}^a +
Q_{\alpha}^a\right)\delta^{4}_{GR} \left(P_{\alpha}^{\dot{a}} +
\tilde{Q}_{\alpha}^{\dot{a}}\right).
\end{eqnarray}

There is also a possibility to obtain completely $SU(4)_R$ covariant
expression for the MHV super form factors. All the solutions of
Grassmannian analyticity constraints (\ref{GrassmannianAnalit}) can
be combined together by introducing the $\mathcal{N}=4$ harmonic
superspace. In such a setup $SU(4)_R$ covariance is manifest in all
expressions. We will give such a formulation of super form factor in
appendix A. However, in unitarity based computations in the MHV
sector for the form factors with operators from the chiral
truncation of $T^{AB}$ such a formulation does not give any
significant computational simplifications. So we will use
$Z^{tree,MHV}_n$ further on instead of harmonic superspace one. In
principle, the situation with the computation of  more complicated
objects (such as the NMHV super form factors, the operators from
different supermultiplets, etc) can be different and harmonic
superspace formulation might be useful.

The projection operators can be written as projectors in terms of
$\eta_A^{'},\eta_A^{''}$. For example, for operators
(\ref{operatorsftomF}) one gets the following projectors in terms of
$\eta_A^{'},\eta_A^{''}$'s, respectively,
\begin{eqnarray}\label{thetaprojectors}
\hat{T}:~1|_{\theta=0} &\rightarrow&
\frac{(\partial^{A}\partial^{B})^{'}
(\partial^{A}\partial^{B})^{''}}{\langle\lambda^{'}\lambda^{''}\rangle^2}|_{\eta^{'}=\eta^{''}=0},\nonumber\\
\hat{T}:~\partial_{\theta}^2|_{\theta=0} &\rightarrow&
\frac{(\partial^{A})^{'}
(\partial^{A})^{''}}{\langle\lambda^{'}\lambda^{''}\rangle}|_{\eta^{'}=\eta^{''}=0},\nonumber\\
\hat{T}:~\partial_{\theta}^4|_{\theta=0}
&\rightarrow&1|_{\eta^{'}=\eta^{''}=0}.
\end{eqnarray}

The algorithm for obtaining a particular component answer from
$Z^{tree,MHV}_n$ can be formulated as follows. One has to apply a
particular projector constructed from $\eta_A^{'},\eta_A^{''}$ to
fix the corresponding operator (like the one  in
(\ref{operatorsftomF})), and then apply a sufficient number of
projectors (\ref{projectors}) on the components in the
$\langle\Omega_n|$ superstate. So the projector operator
$\Pi^{\dot{a}\dot{b}}$ in (\ref{Fsuper-component}) is understood as
the projection of $Z^{tree,MHV}_n$ into the
$\langle\Omega_n|T_{\dot{a}\dot{b}}^{(0)}|0\rangle$ form factor. As
another example, one can obtain an expression for the form factor of
$T_{(2)}^{ab}=Tr(\psi^a_{\alpha}\psi^{b \alpha})$ with external
state consisting of gluons with $+$ polarization, and 2 fermions
$$
\langle g^+_1 \ldots \bar{\psi}_{i~a}^{} \ldots \bar{\psi}_{j~b}^{}
\ldots g^+_n|T_{(2)}^{ab}(q)|0\rangle =
\delta^4(\sum_{k=1}^n\lambda_{\alpha}^k
\tilde{\lambda}_{\dot{\alpha}}^k - q_{\alpha\dot{\alpha}})
\frac{\langle ij\rangle^3}{\langle12\rangle \ldots \langle
n1\rangle}.
$$

For a small number of particles in external state ($n=3$ for example)
one can easily verify this result by direct diagram computation.
Note that to obtain the expected total helicity
$\lambda_{\Sigma}=n-2$ for this form factor one also has to consider
the total helicity of the operator.

\subsection{Soft limit. From form factors to amplitudes}
The form of super form factor $Z^{tree,MHV}_n$
(\ref{T[superFormfactor]}) resembles the superamplitude
(\ref{superamplitude}), and moreover coincides with it in the soft
limit (where supermomentum carried by an operator goes to
$0$)\footnote{We acknowledge the discussion with A. Zhiboedov  who
pointed out the latter fact to us.}, i.e.
\begin{equation} \label{SoftLimit}
Z^{tree,MHV}_n(\{\lambda,\tilde{\lambda},\eta\},0,\{0\}) =
\mathcal{A}_n^{tree,MHV}(\lambda,\tilde{\lambda},\eta).
\end{equation}

The explanation of this fact is the following. It is known that
derivative of  the amplitude $A_n=\langle p_1^{\lambda_1} \ldots
p_n^{\lambda_n}|0\rangle$ (or the correlation function of some local
composite operator of the theory) with respect to the coupling constant
$g$ after the appropriate rescaling of the fields in the Lagrangian
$\mathcal{L}$ is equivalent to insertion of the Lagrangian
 operator with zero momentum in the momentum representation
(similar to duality between the correlation functions and the
amplitudes in \cite{Eden:2010zz})
\begin{equation}
g\frac{\partial A_n}{\partial g} = \langle p_1^{\lambda_1} \ldots
p_n^{\lambda_n}|\mathcal{L}(q=0)|0\rangle.
\end{equation}
On the other hand, the operator $T^{ab}_{(4)}$ of the $\mathcal{T}^{ab}$
which is a chiral part of $T^{AB}$ supermultiplet is equal to the
$\mathcal{N}=4$ Lagrangian written in the chiral form (see
\cite{SuperCor1} for details)
\begin{equation}
T^{ab}_{(4)}=\mathcal{L}^{\mathcal{N}=4}_{chiral} = \int
d^4\theta^{c}_{\alpha} \mathcal{T}^{ab}(q,\theta^{c}).
\end{equation}
Using the fact that the Grassmannian integration is equivalent to
differentiation and the properties (\ref{thetaprojectors}) of
$\hat{T}$ transformation one can write
\begin{equation}
\mathcal{L}^{\mathcal{N}=4}_{chiral}(q) =
\hat{T}[\mathcal{T}^{ab}](q,0),
\end{equation}
which gives
\begin{equation}\label{cojectureAmpl-FF}
Z^{MHV}_n(\{\lambda,\tilde{\lambda},\eta\},0,\{0\})=\hat{T}[\mathcal{F}_n^{MHV}]
(\{\lambda,\tilde{\lambda},\eta\},0,\{0\}) = g\frac{\partial
\mathcal{A}_n^{MHV}(\lambda,\tilde{\lambda},\eta)}{\partial g}.
\end{equation}
This leads to the following conjecture: \emph{the MHV superstate
super form factor with zero operator supermomentum is given by the
logarithmic derivative of the MHV superamplitude with respect to the
coupling constant.}

We will verify this relation at one loop in the MHV sector by direct
computation further in the text. Note also that in the soft limit
the action of $\bar{Q}_{\dot{\alpha}\dot{a}}$ on (\ref{anomaly})
gives 0, as it should be for the superamplitude. The conjectured
relation, however, may be inconsistent for the
$\mbox{N}^k\mbox{MHV}$ sector due to the absence of a smooth limit
for the $\mbox{N}^k\mbox{MHV}$ form factors. Indeed it is known that
NMHV amplitudes contain multiparticle poles. The NMHV form factors
likely share the same property. For example,  using the BCFW
recursion relations one can obtain the four-point NMHV form factor
\cite{Sasha}
$$
\langle \phi^{AB}_1g^+_2g^{-}_3\phi^{AB}_4|T^{(0)}_{AB}|0\rangle =
\frac{\langle13\rangle[24]}{s_{23}\langle12\rangle[34]}+
\frac{\langle34\rangle[24][24]}{s_{23}s_{234}[34]} +
\frac{\langle13\rangle[12]\langle13\rangle}{s_{23}s_{123}\langle12\rangle},
$$
where $s_{ij}=(p_i+p_j)^2,s_{ijk}=(p_i+p_j+p_k)^2$. One observes
that in the limit $q\rightarrow0$ the last two terms in the latter
expression become singular due to the presence of $s_{123}$ and
$s_{234}$ in the denominators.

\section{Reduction procedure for dual pseudoconformal scalar
integrals}

The $D=4$ unitarity based method proved to be an effective tool for
the loop computations
\cite{Bern1,4-loop4gluon,6gloonMHV,5gluonMHVchazo}. Despite the fact
that any ansatz obtained by the $D=4$ unitary-based method should be
confirmed by some direct $D$-dimensional computation or additional
information, the $D=4$ methods can offer the crucial guidance for
constructing a full $D$-dimensional answer \cite{SuperSums}. In the
$\mathcal{N}=4$ SYM it is known that the $D=4$ unitary-based methods
gives a correct answer up to $O(\epsilon)$ at the one-loop order for
the arbitrary number of external legs. It was also observed that it
captures the  correct answer for the four-point amplitude up to four
loops \cite{4-loop4gluon} and for the five-point amplitude up to two
loops \cite{5gluonMHVchazo} for the even part (i.e. part which does
not contain $\gamma^5$) of the amplitude.

In a general (non)supersymmetric theory there is an unknown basis of
scalar integrals beyond one loop. However, for the $\mathcal{N}=4$
SYM theory the situation is different. It was observed
\cite{dualKorch,dual} that the amplitudes in the $\mathcal{N}=4$ SYM
possess a new type of symmetry - the dual conformal symmetry. The
reflection of this symmetry on the level of scalar integrals is that
all scalar integrals entering into the final answer should be
pseudoconformal integrals in momentum space\footnote{In addition to
dual pseudoconformal integrals, if dimensional reduction is used,
there are contributions from the integrals over the $\epsilon$-part
of the loop momenta. Such integrals are not dual conformal; however,
for the MHV sector they likely cancel among themselves if one
considers the BDS exponent \cite{6gloonMHV}.}. This, in fact,
defines the basis of scalar integrals for the $\mathcal{N}=4$ SYM
theory.

In the case of amplitudes the $N_c \rightarrow \infty$ limit
coincides with topologically planar scalar integrals while for form
factors this is not always true. One encounters contributions of the
same order in $N_c$ which contain planar and non-planar (from
topological point of view) scalar integrals. This can be seen, for
example, in studying of the Sudakov form factor $F_2=\langle
\phi^{AB}_1\phi^{AB}_2|T^{(0)}_{AB}|0\rangle $ \cite{BKV_FormFN=1}
where the planar and non-planar diagrams contributing to the answer
at the second order of perturbation theory are of the same order in
$N_c$ (as well as for form factors $Z^{MHV}_n$ discussed
above\footnote{We would like to thank G. Yang for pointing out this
to us.}). In the case of form factors for more general operators
1/2-BPS supermultiplet $Tr \Phi^n$, for example, the mixing between
planar and non-planar scalar integrals of the same order in $N_c$
happens at the $n$-th order. We will split the ratio
$Z^{(m),MHV}_3/Z^{tree,MHV}_3$ into planar and non-planar parts in
the sense of topology for scalar integrals and \emph{will discuss
the planar part only}.

The scalar integrals encountering in loop computations of the  form
factors of the half-BPS operators are not in general pseudo dual
conformal invariant (see \cite{BKV_FormFN=1,FormFMHV}), so one may
think that there are no restrictions for the basis of scalar
integrals apart from the requirement of the UV finiteness. However,
the computations based on the  $\mathcal{N}=1$ coordinate superspace
suggest \cite{BKV_FormFN=1} that the basis of scalar integrals for
the form factors of the half-BPS operators can be obtained from that
of the
 MHV amplitudes applying the reduction procedure
discussed in \cite{BKV_FormFN=1} which is based on the calculation
of triangle and box-type ladder diagrams in \cite{Usyukina:1992jd}.
This gives us the conjectured basis of scalar integrals for the MHV
form factor with $n$ legs at one and two loops. The coefficients of
these integrals are fixed by $D=4$ two-particle iterated cuts, much
in the spirit of the five- and six-point two-loop computations for
the MHV amplitudes.

Let us consider the form factor of the form
$F_n=\langle\Omega_n|\mathcal{O}^{(m)}_{AB}|0\rangle$, where $n\geq
m$ and introduce the notion of the number of "interacting fields"
$V$ for the form factor
\begin{equation}
V=m_{in}+m_{fin},
\end{equation}
as the number of fields participating in the Wick contractions from
the initial operator $\mathcal{O}^{(m)}_{AB}$ and
 from the interaction Lagrangian, $m_{in}$, plus the number of fields participating in the Wick
contractions from
the external super state $\langle\Omega_n|$  and the interaction
Lagrangian, $m_{fin}$ .

\begin{figure}[h]
 \begin{center}
 \leavevmode
  \epsfxsize=10cm
 \epsffile{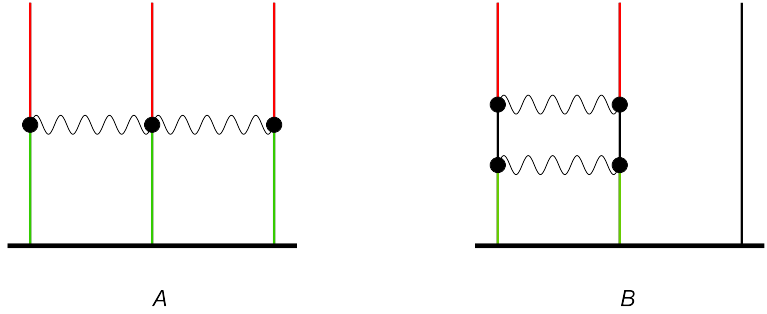}
 \end{center}\vspace{-0.2cm}
 \caption{Some coordinate superspace Feynman diagrams for the
$F_3=\langle \phi_1^{AB} \phi_2^{AB}
\phi_3^{AB}|\mathcal{O}^{(3)}_{AB}|0\rangle$ form factor in two loop
order. The number of red lines is equal to $m_{in}$, the number of
green lines equal to $m_{fin}$. The straight lines correspond to the
chiral propagators $\langle \bar{\Phi}_I^a\Phi_J^b \rangle$, the
wavy lines correspond to the vector  propagator $\langle V^aV^b
\rangle$ of $\mathcal{N}=1$ superfields \cite{BKV_FormFN=1}. The
lower bold line represents the insertion of the corresponding
operator $\mathcal{O}^{(3)}_{AB}$.}\label{fig2}
 \end{figure}

For example, the form factor $F_3=\langle
\phi_1^{AB}\phi_2^{AB}g^+_3|\mathcal{O}^{(2)}_{AB}|0\rangle$ has
$V=2+3,~m_{in}=2,m_{fin}=3$ in any order of PT, and the form factor
$F_n=\langle \phi_1^{AB} \ldots
\phi_n^{AB}|\mathcal{O}^{(n)}_{AB}|0\rangle$ in the  first order of PT
$V=2+2$, and at second order of PT $V$ can take two possible values
$V=2+2$ and $V=3+3$. This  can be easily seen from examples of
$\mathcal{N}=1$ coordinate superspace Feynman diagrams for the
$F_3=\langle \phi_1^{AB} \phi_2^{AB}
\phi_3^{AB}|\mathcal{O}^{(3)}_{AB}|0\rangle$ form factor from
\cite{BKV_FormFN=1}  presented in Fig.
\ref{fig2}. In fact, the definition of $V$ is needed to take into
account the factorized contribution when the number of particles in the
external state $|\Omega_n\rangle$ is equal to the number of fields
in the operator $\mathcal{O}^{(n)}$.

\emph{Conjecture}. The form factors
$F_n=\langle\Omega_n|\mathcal{O}^{(m)}_{AB}|0\rangle$ in the $l$'th
order of PT are given by scalar integrals which are obtained by
reduction procedure from the scalar pseudoconformal integrals
appearing at the $l$-loop in the $V$-point MHV amplitudes. This
procedure involves the shrinking of $m_{in}-1$ propagators which connect
$m_{in}$ external legs standing in a row for the corresponding
amplitude.
If the $m_{in}$ neighboring
momenta are attached to the same vertex, then there are no
contractions. The shrinking of a propagator to a point can also be
 understood in dual variables as taking the $m_{in}-1$ external legs
(more accurately points) to infinity \cite{BKV_FormFN=1}. The
integrals containing the UV divergent subgraphs (bubbles) are not
taken into account. Heuristically, this rule can be understood as
the consequence of the appearance of a new type of the "effective
vertex" in the unitarity cuts - the form factor, which glues several
momenta together.

\begin{figure}[h]
 \begin{center}
 \leavevmode
  \epsfxsize=12cm
 \epsffile{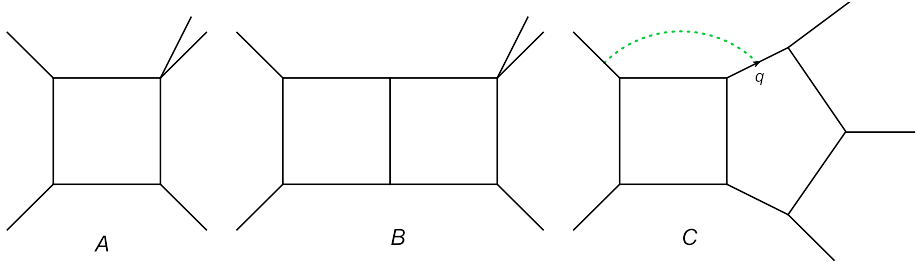}
 \end{center}\vspace{-0.2cm}
 \caption{The basis of scalar integrals through $O(\epsilon)$
 at one (A) and two (B,C) loops for the even part of the five-point MHV amplitude.
 Green ark corresponds to the presence of the
 numerator.}\label{fig1}
 \end{figure}
As an example of the above mentioned procedure one can apply it to
find the basis of scalar integrals for one- and two-loop three-point
super form factor from the corresponding one- and two-loop five-point MHV amplitude.
 At one loop one can choose as the basis for the integrals
 the combination of boxes (see
Fig. \ref{fig1} A), and at two loops - the combination of diagrams
shown in Fig. \ref{fig1} B and C. The result of the
application of the reduction procedure is presented in Fig. \ref{fig3}
(we consider only the even part since the odd part gives no contribution for
three-point kinematics of the form factor). Hence, one expects  to
obtain the answer for three-point MHV form factor  in terms of the
following set of scalar integrals:
\begin{figure}[h]
 \begin{center}
 \leavevmode
  \epsfxsize=13cm
 \epsffile{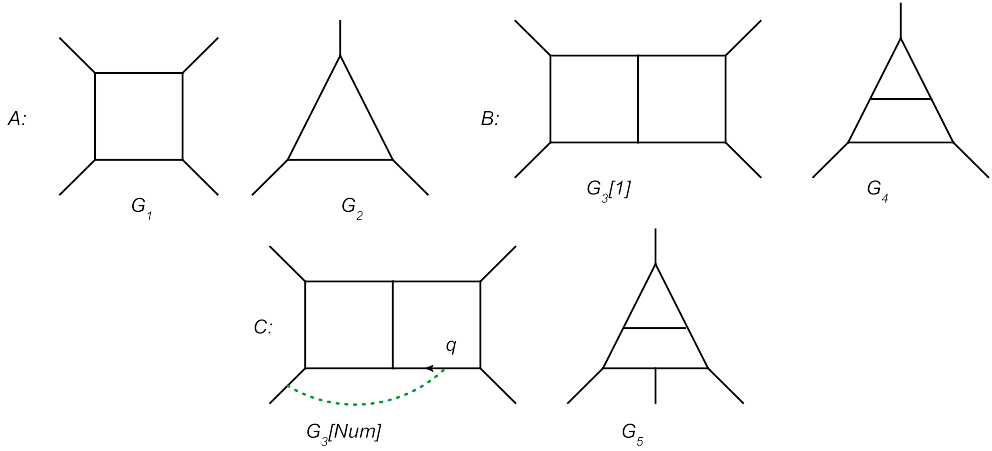}
 \end{center}\vspace{-0.2cm}
 \caption{Scalar integrals obtained by the reduction procedure. We conjecture
 that they form a basis of scalar integrals for the $n=3$-point MHV form
 factor at one (A) and two (B,C) loops.}\label{fig3}
 \end{figure}

$$
\mbox{1-loop:}~\textbf{G}_{1},\textbf{G}_{2},$$
$$\mbox{2-loops:}~\textbf{G}_{3},\textbf{G}_{3}[Num.],\textbf{G}_{4},\textbf{G}_{5}.
$$

\begin{figure}[h]
 \begin{center}
 \leavevmode
  \epsfxsize=7cm
 \epsffile{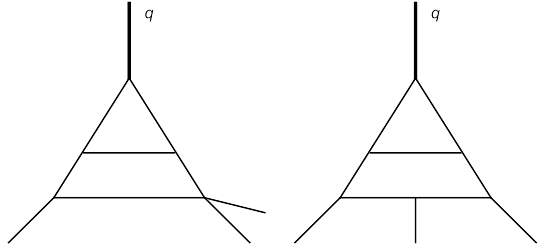}
 \end{center}\vspace{-0.2cm}
 \caption{The configuration of external momenta
 for $\textbf{G}_5$ and $\textbf{G}_4$ scalar integrals.}\label{fig4}
 \end{figure}

All these integrals can be captured by two particle iterated cuts. Note also that the reduction procedure, at least in the
current form, is "blind" to the external momenta configuration, and,
in principle, one has to consider all different combinations.
However, it is natural to assume that the off-shell momenta of the
operator should be attached to the contracted propagator. We use the
pattern which is represented in Fig. \ref{fig4}.

\section{$n$-point MHV super form factor at one loop}
\begin{figure}[h]
 \begin{center}
 \leavevmode
  \epsfxsize=8cm
 \epsffile{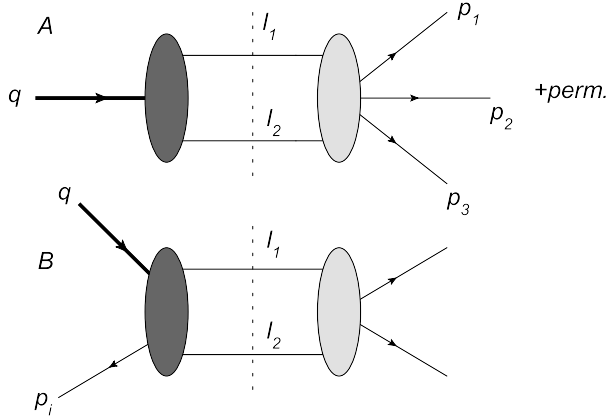}
 \end{center}\vspace{-0.2cm}
 \caption{All two particle iterated cuts for the $n=3$ point MHV form factor at
 one loop. Dark grey vertex corresponds to the MHV tree form factor,
 light grey vertex corresponds to the MHV tree amplitude.}\label{OneLoopFig}
 \end{figure}

Here we will reproduce the results obtained in \cite{FormFMHV} for the
three-point MHV form factors using "superstate - super form factors",
and briefly consider the case for a general $n$-point MHV super form
factor. To obtain the full answer for the one loop MHV super form factor
$Z^{(1),MHV}_3$, one has to consider the following cuts (see Fig.
\ref{OneLoopFig}). Since the operator is color singlet and the external
state is color-ordered, all possible cyclic permutations of external
momenta should be considered \cite{FormFMHV}.
Let us consider first the cut $A$ shown in Fig. \ref{OneLoopFig}
\begin{eqnarray}
disc_{q^2}[Z^{(1),MHV}_3]=\int
dSLIPS^{l_1l_2}_2Z^{tree,MHV}(-l_1,-l_2)
\mathcal{A}_5^{tree,MHV}(1,2,3,l_1,l_2) + \mbox{perm},
\nonumber\\
\end{eqnarray}
where $$q = l_1 + l_2 = p_1 + p_2 + p_3,$$ and $\mbox{perm}$ means
the terms with the exchanged momenta $(1,3)$ by
$(2,1)$ and $(3,2)$. Performing the supersummation  one gets
\begin{eqnarray}\label{1loop(q^2)Cut}
disc_{q^2}[Z^{(1),MHV}_3] = \lambda Z^{tree,MHV}_3\int
dLIPS^{l_1l_2}_2\frac{\langle l_1l_2 \rangle
\langle13\rangle}{\langle 1l_1 \rangle \langle 3 l_2 \rangle} +
\mbox{perm}.
\end{eqnarray}

The integrand is identical to the integrand obtained in
\cite{FormFMHV}. Replacing $dLIPS^{l_1l_2}_2\rightarrow
d^Dl_1/(2\pi)^Dl_1^2l_2^2$, where $l_1$ is an unrestricted loop
momentum, we rearrange the integrand as
\begin{equation}
\frac{\langle l_1l_2 \rangle \langle13\rangle}{\langle
3l_1\rangle\langle
l_21\rangle}=\frac{C_1}{2(l_21)}+\frac{C_3}{2(l_13)}+\frac{D_{13}}{2(l_13)2(l_21)},
\end{equation}
where the notation $p_i\equiv i$, $(p_i+p_j)^2\equiv 2(ij)$ is
used and also
\begin{equation}
C_i=(iq), D_{ij}=2(ik)(kj),i\neq k,j\neq k.
\end{equation}
We see the part of $Z^{(1),MHV}_3$ has a discontinuity in the $q^2$
channel
\begin{eqnarray}\label{MHV1loopanswer}
&&Z^{(1),MHV}_3|_{q^2} = \lambda Z^{tree,MHV}_3 M_3^{(1)},\nonumber\\
&&M_3^{(1)} = \left[ D_{13}\textbf{G}_{1}(1,2,3|(-q)^2)+
C_1\textbf{G}_{2}(1|(-q)^2,s_{23}) \right. \nonumber \\ && \left.
\makebox[2.85em]{} + C_3\textbf{G}_{2}(3|(-q)^2,s_{12}) \right] +
\mbox{perm},
\end{eqnarray}
where $\mbox{perm}$ means the permutation over the external legs
which leads to the doubling of the triangles $\textbf{G}_2$ in the
above expression. The following notation for the integrals was used
$$
\textbf{G}_{i}[Num.](\mbox{massless legs}|(\mbox{massive legs})^2).
$$

The case with the cut $B$ from Fig.
\ref{OneLoopFig} does not bring any new information
\cite{FormFMHV}. Nevertheless, let us consider this cut, since it
introduces some features which will be encountered in the two-loop
calculation. Let us choose $i=3$
\begin{eqnarray}
disc_{(q-3)^2}[Z^{(1),MHV}_3]=\int
dSLIPS^{l_1l_2}_2Z^{tree,MHV}_2(l_2,l_1,3)\mathcal{A}_4^{tree,MHV}(1,2,-l_1,-l_2),
\end{eqnarray}
where $$q - p_3 = l_1 + l_2 = p_1 + p_2.$$

Performing the supersummation and replacing
$dLIPS^{l_1l_2}_2\rightarrow d^Dl_1/(2\pi)^Dl_1^2l_2^2$ we can write
the part of $Z^{(1),MHV}_3$ which has a discontinuity in the
$(q-3)^2$ channel as
\begin{eqnarray}\label{1loop(q-i)Cut}
Z^{(1),MHV}_3|_{(q-3)^2} = \lambda Z^{tree,MHV}_3\int
\frac{d^Dl_1}{(2\pi)^Dl_1^2l_2^2}\hat{R}(3,3),
\end{eqnarray}
here $\hat{R}(a,b)$ is a universal one-loop function \cite{Bern1}
defined as
\begin{equation}
\hat{R}(a,b) = R(b,a+1) + R(b-1,a) - R(a,b) - R(b-1,a+1),
\end{equation}
where
\begin{eqnarray}
R(b,a) &=& 1+\frac{C_b}{2(bl_1)} + \frac{C_a}{2(al_2)} +
\frac{D_{ba}}{2(bl_1)2(al_2)},
\nonumber \\
C_a &=& (a P),D_{ab} = 2(a P)(b P)-P^2(ab).
\end{eqnarray}

In our case $P=q-3$ and it gives
$$
\hat{R}(3,3) = R(3,1) + R(2,3) - R(3,3) - R(2,1),
$$
where $R(3,3) = 1$ \cite{FormFMHV} and $D_{12} = 0$.

So $\hat{R}(3,3)$ takes the form
\begin{equation}\label{R33}
\hat{R}(3,3)=\frac{C_3}{2(3l_1)}+\frac{C_3}{2(3l_2)}+\frac{D_{31}}{4(3l_1)(1l_2)}
+\frac{D_{23}}{4(2l_1)(3l_2)}.
\end{equation}
Using this we can arrange $Z^{(1),MHV}_3|_{(q-3)^2}$ as (we are
dropping the common $\lambda Z^{tree,MHV}_3$ prefactor)
\begin{eqnarray}
D_{13}\textbf{G}_{1}(1,2,3|(-q)^2)+D_{23}\textbf{G}_{1}(2,1,3|(-q)^2)+
2C_3\textbf{G}_{2}(3|(-q)^2,s_{12}).
\end{eqnarray}

All these integrals, as expected, are contained in
(\ref{MHV1loopanswer}). Similar cuts in the $(q-2)^2$- and
$(q-1)^2$- channels should be considered in the same way, so the
full answer for $Z^{(1),MHV}_3$ is given by the following expression
\begin{eqnarray}\label{MHV1loopanswer1}
 Z^{(1),MHV}_3 &=& \lambda Z^{tree,MHV}_3 M_3^{(1)},\\
 M_3^{(1)} &=&
D_{13}\textbf{G}_{1}(1,2,3|(-q)^2)+D_{12}\textbf{G}_{1}(1,3,2|(-q)^2)+D_{23}\textbf{G}_{1}(2,1,3|(-q)^2)
\nonumber\\
&+&2C_1\textbf{G}_{2}(1|(-q)^2,s_{23})+
2C_2\textbf{G}_{2}(2|(-q)^2,s_{13}) +
2C_3\textbf{G}_{2}(3|(-q)^2,s_{12}) .\nonumber
\end{eqnarray}

The case with $n$ legs is treated in a similar way. For the cuts in
$q^2$- and $(q-i)^2$- channels one gets similar contributions as
in \cite{FormFMHV}.
Apart from the previous cuts there is also
another new kinematic channel in $s_{a+1,b-1}=(p_{a+1}+ \ldots
+p_{b-1})^2, a,b=1, \ldots, n; a \neq b$. The result in this channel
is
\begin{eqnarray}\label{1loopGeneralCut}
Z^{(1),MHV}_n|_{s_{a+1,b-1}} = \lambda Z^{tree,MHV}_3\int
\frac{d^Dl_1}{(2\pi)^Dl_1^2l_2^2}\hat{R}(a,b),
\end{eqnarray}
which coincides with the results of \cite{FormFMHV} and gives all
the possible combination of two mass-easy and one mass-easy boxes in
addition to triangles in the $q^2$ and $(q-i)^2$ channels.

Let us discuss several soft limits. First we consider
$q\rightarrow0$ for $n\geq4$ (in the cases when $n=2,3$ the
kinematics becomes degenerate). At the tree-level the connection is
given by (\ref{SoftLimit}), so it is interesting to verify whether
such a relation holds at the one-loop level. All the contributions
in the $q^2$-channel and $(q-i)^2$-channel vanish in this limit,
because all the coefficients such as $C_i$ and $D_{ij}$ contain at
least a linear dependence on $q$. The remaining contributions coming
from the $s_{a+1,b-1}, a \neq b$ channels reproduce exactly the one-loop
$n$-point MHV amplitude \cite{Bern1}
\begin{eqnarray}
Z^{(1),MHV}_n|_{s_{a+1,b-1}}^{q=0} = \mathcal{A}^{(1),MHV}_n|_{s_{a+1,b-1}} = \lambda \mathcal{A}^{tree,MHV}_n
\int\frac{d^Dl_1}{(2\pi)^Dl_1^2l_2^2}\hat{R}(a,b).
\end{eqnarray}

Let us consider also the three-point form factor in the limit where
the momentum of the third external particle becomes soft. We would
expect that in this limit the ratio $Z^{(1),MHV}_3/Z^{tree,MHV}_3$
would give us the one-loop planar part of the $M^{(1),planar}_2$
form factor -- the one-loop Sudakov form factor
\cite{vanNeerven:1985ja,BKV_FormFN=1}. Indeed, one can see that in
this limit all $D_{ij}$ in (\ref{MHV1loopanswer}) vanish and
$\textbf{G}_{2}$ integrals give us just the necessary combination
\begin{equation}
M^{(1),planar}_2 = 2 s_{12} \textbf{G}_{2}(1,2|s_{12}).
\end{equation}
We would expect also that similar behaviour holds also at the
higher-loop level. For more details on the soft and collinear limits
see Appendix B.

\section{Ansatz for three-point MHV super form factor at two loops}
\begin{figure}[h]
 \begin{center}
 \leavevmode
  \epsfxsize=8cm
 \epsffile{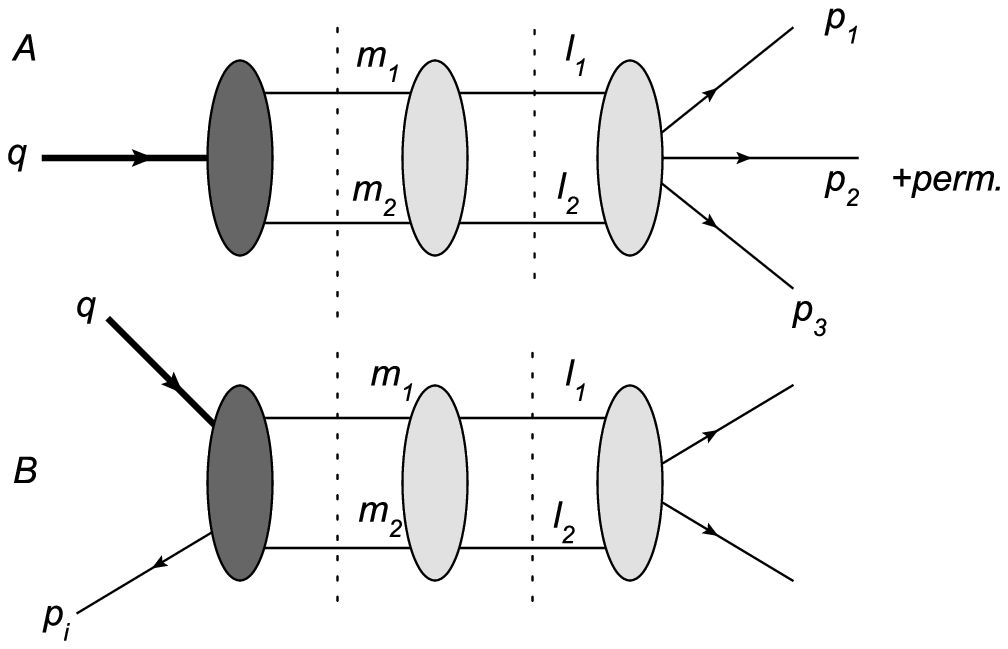}
 \end{center}\vspace{-0.2cm}
 \caption{All two-particle iterated cuts for the three-point MHV form factor at
 two loops. Dark grey vertex corresponds to he MHV tree form factor,
 while the light grey vertex to the  MHV tree amplitude.} \label{2loop3point}
 \end{figure}

We consider here the  $D=4$ two-particle iterated cuts for
three-point super form factors at two loops based on two-particle
iterated cuts, and suggest the ansatz for the three-point super form
factors at two loops based on the assumed basis of scalar integrals
(which is obtained by  reduction from the basis of the
scalar pseudoconformal integrals for the amplitudes). The needed unitary cuts
 for the three-point form factor are shown in Fig.
\ref{2loop3point}.

Let us consider the cut $A$ first.  Using the
momentum conservation relation
$$
q = l_1 + l_2 = m_1 + m_2 = p_1 + p_2 + p_3,
$$
performing the supersummation and replacing $dLIPS^{l_1l_2m_1m_2}_4\rightarrow
d^Dl_1d^Dm_1/(2\pi)^{2D}l_1^2l_2^2m_1^2m_2^2$ where $l_1,m_1$ are
unrestricted loop momenta one writes the part of $Z^{(2),MHV}_3$ which
has a double discontinuity (i.e. discontinuity of discontinuity) in the
$q^2$-channel as
\begin{eqnarray}
\lambda^2 Z^{tree,MHV}_3\int
\frac{d^Dl_1d^Dm_1}{(2\pi)^{D}(2\pi)^{D}}\frac{1}{l_1^2l_2^2m_1^2m_2^2}\frac{\langle31\rangle\langle
l_1l_2 \rangle^2\langle m_1m_2\rangle}{\langle 1l_1 \rangle \langle
3l_2 \rangle \langle l_1 m_1 \rangle \langle l_2 m_2 \rangle} +
\mbox{perm},
\end{eqnarray}
which is combined into (we are dropping the common $\lambda^2 Z^{tree,MHV}_3$
prefactor)
\begin{eqnarray}
&& q^2 \left[ D_{13}\textbf{G}_{5}(1,2,3|(-q)^2) +
2C_1\textbf{G}_{4}(1|(-q)^2,s_{23}) \right. \nonumber \\ && \left.
\makebox[3em]{} + 2C_3\textbf{G}_{4}(3|(-q)^2,s_{12}) \right] +
\mbox{perm},
\end{eqnarray}
from which we see that all the conjectured integrals except for some
double boxes were captured by the cut $A$.

Let us consider now the cut $B$ and
take $i=3$ (for any other $i$ the procedure is the same). Taking
into account the momentum conservation laws which are
$$
q-p_3 = l_1 + l_2 = m_1 + m_2 = p_1 + p_2,
$$
and performing the supersummation one can rewrite the  part of $Z^{(2),MHV}_3$
which has the double discontinuity in the $(q-3)^2$-channel as
\begin{eqnarray}
\lambda^2 Z^{tree,MHV}_3 (12) \int\frac{d^Dm_1}{(2 \pi)^D
m_1^2m^2_2} \mbox{Box}(2,m_1)~\hat{R}(3,3),
\end{eqnarray}
where $$\mbox{Box}(x,y) = \int \frac{d^Dl_1}{(2 \pi)^D
l_1^2l_2^2}\frac{1}{(xl_2)(yl_2)}.$$

Using the properties of $\hat{R}(3,3)$ discussed in the previous
section one gets (again the common $(12) \lambda^2 Z^{tree,MHV}_3$ factor is
dropped) $$ \textbf{G}_{3}[C_3(2+m_1)^2+D_{13}](1,2,3|(-q)^2)
+\textbf{G}_{3}[C_3(1+m_1)^2+D_{32}](1,2,3|(-q)^2),$$ thus obtaining
the expected double boxes.

The cuts in the $(q-1)^2$- and $(q-2)^2$- channels give us the same
boxes with the numerators
$$
(23)C_1(3+m_1)^2+(23)D_{21},~(23)C_1(2+m_1)^2+(23)D_{13},$$
$$(13)C_2(3+m_1)^2+(13)D_{12},~(13)C_2(1+m_1)^2+(13)D_{23}.
$$
Combining all contributions together we can write the following
ansatz for the three-point MHV two-loop form factor
$$Z^{(2),MHV}_3 = \lambda^2 Z^{tree,MHV}_3 M_3^{(2)},$$
with
\begin{eqnarray}\label{MHV1loopanswer2}
&& M_3^{(2)} = \left[ q^2D_{13} \textbf{G}_{5}(1,2,3|(-q)^2) +
2q^2C_1 \textbf{G}_{4}(1|(-q)^2,s_{23}) \right. \nonumber \\ &&
\left. \makebox[-2em]{} + 2q^2C_3 \textbf{G}_{4}(3|(-q)^2,s_{12}) +
s_{12}\textbf{G}_{3}[C_3(2+m_1)^2+D_{13}](1,2,3|(-q)^2) \right. \nonumber \\
&& \left. \makebox[2em]{} +
s_{12}\textbf{G}_{3}[C_3(1+m_1)^2+D_{32}](1,2,3|(-q)^2) \right] +
\mbox{perm}.
\end{eqnarray}

We see that this expression in the limit when momenta for
the third external leg goes to zero  gives the
expected expression \cite{vanNeerven:1985ja,BKV_FormFN=1}
\begin{equation}
\frac{Z^{(2),MHV}_3}{Z^{tree,MHV}_3}|_{3\rightarrow0}=M^{(2),planar}_2
= 4 s_{12}^2 \textbf{G}_{4}(1,2|s_{12}).
\end{equation}

This ansatz for the three-point MHV form factor in the second order
of PT is based on the conjecture that the basis of scalar integrals
for the form factors is obtained from the basis of scalar integrals
for particular (MHV) amplitudes, using the reduction procedure first
introduced in \cite{BKV_FormFN=1} and studied here for the purposes
of our calculations. It still needs verification concerning the
factorization properties, collinear and soft limits, as well as the
computation none planar (in the sense of the topology of the scalar
integrals) contribution. We are going to address these questions in
upcoming publications.

\section{Discussion}
In this paper, the systematic study of form factors in the
$\mathcal{N}=4$ SYM theory is performed. Initially, they were studied
long time ago in \cite{vanNeerven:1985ja}, \cite{Perturbiner} and
revived recently first in the strong coupling \cite{Maldacena:2010kp}
and then in the weak coupling regime in \cite{BKV_FormFN=1}, where the
$\mathcal{N}=1$ superspace calculation was used, and in
\cite{FormFMHV}, where the unitary-based technique for constructing
the one-loop form factors was applied. In this paper the manifestly
$\mathcal{N}=4$ SUSY covariant answer is obtained for the form
factor of the operator belonging to the stress-tensor energy
supermultiplet, and the completely $\mathcal{N}=4$ SUSY covariant
unitary-based technique for constructing the loop form factors is
presented.

The central result of this paper is the formula for the MHV tree
level supersymmetric form factor (\ref{T[superFormfactor]}) for the
operators belonging to the stress-tensor energy supermultiplet (its
chiral truncation) and its connection to the superamplitude given by
(\ref{cojectureAmpl-FF}). This hypothesis links the  super form
factor with super momentum equal to zero and the derivative of the
superamplitude. The latter conjecture has been verified at the tree-
and one-loop level for any MHV $n$-point ($n\geq4$) form factor
involving the stress-tensor energy supermultiplet. Apart from this
the collinear and soft limits of form factors were considered which
are very similar to the corresponding limits for the amplitudes. In
the case of form factors, there are two different soft limits: the
first one is applied to one of the external on-shell legs bringing
the form factor with $n$ external legs to the form factor with $n-1$
legs. The other one is the soft limit applied to the momentum of the
initial operator which brings to the amplitude.

The two-loop result for the three-point super form factor is
constructed based on two-particle iterative cuts and the hypothesis
that the basis of scalar integrals is known. It was known for a long
time that the two-point form factor in the $\mathcal{N}=4$ SYM
theory has a very simple structure \cite{vanNeerven:1985ja}: the
anomalous cusp and collinear dimensions determines the divergent
part while the finite part is trivial (constant, and is conjectured
to be completely determined by the one loop result in some
renormalization scheme \cite{vanNeerven:1985ja}) and exponentiates.
This is very similar to the behavior of the four- and five-point MHV
amplitudes where the finite part is also simple and exponentiates.
The latter can be understood as a consequence of the dual conformal
invariance. It is natural to consider the two-point form factor as an
analog of the four-point amplitude \cite{bds05} and three-point form
factor corresponds to the five-point one \cite{5gluonMHVchazo}. Then
it is interesting to verify if the two-loop answer for the
three-point form factor obtained here has a  simple structure and
exponentiate. This requires the calculation of the basic scalar
integrals. If it is really the case it would be the indication that
the form factors are also governed by some hidden symmetry.

The basis of scalar integrals appearing in form factor calculation
deserves a separate comment. This basis is not  arbitrary but
can be obtained by the reduction procedure from the pseudoconformal
integrals which appear in the amplitude calculations. This procedure was
observed earlier in \cite{BKV_FormFN=1} where a completely different
approach was used for studying the form factors, namely, the
$\mathcal{N}=1$ superspace technique. The main idea of this
reduction procedure is the following: to get the scalar integrals
for form factors, one should shrink some of the propagators in scalar
pseudoconformal integrals forming  the basis of
integrals for the amplitudes.

\section*{Acknowledgements}

We would like to thank A. A. Gorsky, T. McLoughlin, I. B. Samsonov,
E. A. Ivanov and V. A. Smirnov for valuable discussions. We
especially would like to thank A. V. Zhiboedov for kindly sharing
some of his results \cite{Sasha} with us and enlightening
discussions. Financial support from RFBR grant \# 08-02-00856 and
the Ministry of Education and Science of the Russian Federation
grant \# 1027.2008.2 is kindly acknowledged. LB thanks the Dynasty
Foundation for financial support.

\

{\bf Note added}: while finishing writing the paper we became aware
of the paper which is closely connected to the subject studied here
and has an overlap in deriving the super form factor
\cite{Brandhuber:2011tv}.

\appendix

\section{$\mathcal{N}=4$ harmonic superspace}
We  discuss here the reformulation of (\ref{T[superFormfactor]})
in the $\mathcal{N}=4$ harmonic superspace. Our discussion is based mostly on
section 3 of \cite{SuperCor1}. The $\mathcal{N}=4$ harmonic superspace is
obtained by adding additional bosonic coordinates (harmonic
variables) to the $\mathcal{N}=4$ coordinate superspace or on-shell momentum
superspace. These additional bosonic coordinates parameterize the coset
$$
\frac{SU(4)}{SU(2) \times SU(2) \times U(1)}
$$
and carry the  $SU(4)$ index $A$, two copies of $SU(2)$ indices
$a, \dot{a}$ and $U(1)$ charge $\pm$
$$
(u^{+a}_{A},~u^{-\dot{a}}_A).
$$

Using these variables one presents all the Grassmannian objects with
$SU(4)_R$ indices. For example, for Grassmannian coordinates in the
original $\mathcal{N}=4$ coordinate superspace
\begin{equation}
\theta^{+a}_{\alpha}=u^{+a}_{A}\theta^A_{\alpha},~
\theta^{-\dot{a}}_{\alpha}=u^{-\dot{a}}_{A}\theta^A_{\alpha},
\end{equation}
and in the opposite direction
\begin{equation}
\theta^A_{\alpha}=\theta^{+a}_{\alpha}u_{+a}^{A}+\theta^{-\dot{a}}_{\alpha}\bar{u}_{+a}^{A}.
\end{equation}

The same can be done with supercharges, projection operators from
(\ref{projectors}) etc.. Note that harmonic variable projection
leaves helicity properties of the objects unmodified. Also, similar
projections can be performed for Grassmannian coordinates $\eta^A$ and
supercharges $q_{\alpha}^A,~\bar{q}_{\dot{\alpha}A}$ of on-shell
momentum superspace.

So the $\mathcal{N}=4$ harmonic superspace is parameterized with the
following set of coordinates
\begin{eqnarray}
\mbox{$\mathcal{N}=4$ harmonic superspace}&=&\{x^{\alpha\dot{\alpha}},
~\theta^{+a}_{\alpha},\theta^{-\dot{a}}_{\alpha},
~\bar{\theta}_{\dot{\alpha}}^{+a},\bar{\theta}_{\dot{\alpha}}^{-\dot{a}},u
\}\nonumber\\
\mbox{or}&&\{\lambda_{\alpha},\tilde{\lambda}_{\dot{\alpha}},~\eta^{+a},\eta^{-\dot{a}},~u\}.
\end{eqnarray}
Using $u$ harmonic variables one can project the $W^{AB}$ superfield as
$$
W^{AB}\rightarrow W^{AB}u^{+a}_{A}u^{+b}_{B}=\epsilon^{ab}W^{++},
$$
where $\epsilon^{ab}$ is an $SU(2)$ totally antisymmetric tensor
and the Grassmannian analyticity conditions (\ref{GrassmannianAnalit}) such
that\footnote{Strictly speaking this is true only in the free theory
($g=0$), in the interacting theory one has to replace
$D_{\alpha}^A,\bar{D}_{\dot{\alpha}}^A$ by their gauge covariant
analogs, which contain superconnection, but the final result is the
same \cite{SuperCor1}.}
$$
D^{\alpha}_{-\dot{a}}W^{++}=0,~ \bar{D}^{\dot{\alpha}}_{+a}W^{++}=0.
$$

Thus, the superfield $W^{++}$ contains the dependence on half of the Grassmannian variables
$\theta$'s and $\bar{\theta}$'s.
$$
W^{++}=W^{++}(x, ~\theta^{+a}_{\alpha},
\bar{\theta}_{\dot{\alpha}}^{-\dot{a}},u),
$$

Performing the expansion of $W^{++}$   in $u$ \emph{all} the
projections like (\ref{prrojection2}) in $SU(4)_R$ covariant
fashion can be obtained. This is the main purpose of introduction of the harmonic superspace.
The component expansion of $W^{++}$ in $\theta$'s and
$\bar{\theta}$'s can be found in \cite{SuperCor1}. The lowest
component of the $W^{++}$ expansion is
$$
W^{++}(x,0,0,u)=\phi^{++},~\phi^{++}=\frac{1}{2}\epsilon_{ab}u^{+a}_{A}u^{+b}_{B}\phi^{AB},
$$
where according to \cite{SuperCor1}
\begin{equation}\label{QS++=0}
Q_{\alpha}^{-\dot{a}}\phi^{++}=0.
\end{equation}

Using this condition we can write in analogy with
(\ref{symmetryConditionsForF}) that the form factor
\begin{equation}
F_{n}(\{\lambda,\tilde{\lambda},\eta\},q,u)=\langle\Omega_n|Tr(\phi^{++}\phi^{++})|0\rangle,
\end{equation}
satisfies the following condition:
$$
q_{\alpha}^{-\dot{a}}F_{n}(\{\lambda,\tilde{\lambda},\eta\},q,u)=0,
$$
where $q_{\alpha}^{-\dot{a}}$ is the projected supercharge in the on
shell momentum superspace representation
$q_{\alpha}^{-\dot{a}}=u^{-\dot{a}}_Aq_{\alpha}^{A}$. We see that
$$
F_{n} \sim \delta^{-4}(q^{\dot{a}}_{\alpha})(\ldots),
$$
where $\delta^{-4}$ is the Grassmannian delta function; $\delta^{\pm4}$ are
defined as
\begin{equation}
\delta^{\pm4}(q^{a/\dot{a}}_{\alpha})=\sum_{i,j=1}^{n}\prod_{a/\dot{a},b/\dot{b}=1}^2\langle
ij \rangle \eta^{\pm a/\dot{a}}_i\eta^{\pm b/\dot{b}}_j.
\end{equation}

Since the $u$ projections do not change the helicity structure of
Grassmannian variables, using the previous arguments we can write at the
tree level
$$
F_{n}^{tree,MHV}(\{\lambda,\tilde{\lambda},\eta\},q,u) \sim
\frac{\delta^{-4}(q^{a}_{\alpha})}{\langle 12 \rangle \ldots \langle
n1 \rangle}.
$$
This expression is an $SU(4)_R$ invariant analog of
(\ref{Fsuper-component}), which contains all form factors of the form
$\langle\Omega_n|T^{AB}_{(0)}|0\rangle$.

The superfield $W^{++}$ can also be used to combine all the components of the
chiral sector of $T^{AB}$. Letting $\bar{\theta}$'s in $W^{++}$ to
0 the superfield $\mathcal{T}$ \cite{SuperCor1}
$$
\mathcal{T}(x,~\theta^{+a}_{\alpha},u)=Tr(W^{++}W^{++})(x,~\theta^{+a}_{\alpha},u),
$$
contains all projections like $\mathcal{T}^{ab}$. We can write
$\mathcal{T}$ as
$$
\mathcal{T}(x,~\theta^{+a}_{\alpha},u)=e^{Q^{+a}_{\alpha}\theta_{+a}^{\alpha}}\mathcal{T}(x,0,u),
$$
so that for "super state -super form factor" we have
\begin{equation}
\mathcal{F}_n(\{\lambda,\tilde{\lambda},\eta\},q,u,\theta^{+a}_{\alpha})
=\langle\Omega_n|\mathcal{T}(\theta^{+a}_{\alpha})|0\rangle=e^{q^{+a}_{\alpha}\theta_{+a}^{\alpha}}F_n(\{\lambda,\tilde{\lambda},\eta\},q,u)
.
\end{equation}

One can define an analog of $\hat{T}$ transformation for
$\theta_{+a}^{\alpha}$, so that for the MHV part of $\mathcal{F}_n$
at tree level one can write:
\begin{equation}
\hat{T}[\mathcal{F}_n^{tree,MHV}]\sim
\frac{\delta^{+4}(q^{a}_{\alpha}+\lambda^{'}_{\alpha}\eta^{'a}+\lambda^{''}_{\alpha}\eta^{''a})\delta^{-4}(q^{\dot{a}}_{\alpha})}{\langle
12 \rangle \ldots \langle n1 \rangle}.
\end{equation}
This expression looks just like (\ref{T[superFormfactor]}), but now
both the Grassmannian delta functions $\delta^4$  are $SU(4)_R$ covariant.
One can write also the MHV part of a superamplitude  in a similar manner.
Projecting the condition of superamplitude invariance under
$q^{A}_{\alpha}$ supersymmetry transformations we have
$$
q^{A}_{\alpha}\mathcal{A}_n^{tree,MHV}=0\rightarrow
(q^{+a}_{\alpha}+q^{-\dot{a}}_{\alpha}) \mathcal{A}_n^{tree,MHV}=0,
$$
and taking into account that the helicity properties of projected
supercharges are not modified we get
\begin{equation}
\mathcal{A}_n^{tree,MHV}\sim
\frac{\delta^{+4}(q^{a}_{\alpha})\delta^{-4}(q^{\dot{a}}_{\alpha})}{\langle
12 \rangle \ldots \langle n1 \rangle}.
\end{equation}
Now both $\mathcal{A}_n^{tree,MHV}$ and
$\hat{T}[\mathcal{F}_n^{tree,MHV}]$ are $SU(4)_R$ invariant and
one can use them in unitarity based computations, where super summation
should be performed separately for $\delta^{+4}$ and $\delta^{-4}$.

\section{Collinear and soft limits}
Here we gather together the results for collinear and soft limits
for the  form factors at the tree- and one-loop levels. In the limit
when momenta $i$ of external particle becomes soft at the tree level
it is easy to see that similarly to the soft limit in the amplitudes
\cite{Bern1} one has
$$
Z_n^{tree,MHV}\stackrel{i\rightarrow0} \rightarrow
\mbox{Soft}^{tree}(a,i,b)Z_{n-1}^{tree,MHV},~n\geq3,
$$
where the $\mbox{Soft}^{tree}(a,i,b)$ is the "eikonal" factor
\begin{equation}
\mbox{Soft}(a,i,b)=\frac{\langle  ab\rangle}{\langle
ai\rangle\langle  ib\rangle},
\end{equation}
with $a$ and $b$ being the momenta of color-ordered neighbors of the
particle with momentum $i$.

At the one-loop level taking the soft limit in expressions
(\ref{1loop(q^2)Cut}), (\ref{1loop(q-i)Cut}), and
(\ref{1loopGeneralCut}) one obtains
$$
\frac{Z^{(1),MHV}_n}{Z^{tree,MHV}_n}|_{i\rightarrow0}\sim
M^{(1)}_{n-1},~n\geq4.
$$

$Z^{(m),MHV}_3$ in general contains the planar and none planar
scalar integrals. Taking the limit when the momentum of the $i$-th
external leg goes to zero, one gets $Z^{(m),MHV}_2$ which is the
Sudakov form factor. The color structure in this case is trivial;
one has $Tr(T^aT^b) = \frac 12 \delta^{ab}$ and the planar and
non-planar diagrams are of the same order in $N_c$, as was explained
earlier. One would expect that relation between $Z^{(m),MHV}_3$ and
the part of $Z^{(m),MHV}_2$, which contains the planar and none
planar diagrams will hold separately
\begin{equation}
\frac{Z^{(m),MHV,planar/none~planar}_3}{Z^{tree,MHV}_3}|_{3\rightarrow0}
\sim M^{(m),planar/none~planar}_2,
\end{equation}
at one and two loops.

Taking the soft limit when the (super)momenta of an operator becomes
soft one derives
\begin{equation}
Z^{tree/(1),MHV}_n(\{\lambda,\tilde{\lambda},\eta\},0,\{0\})=\mathcal{A}_n^{tree/(1),MHV}(\lambda,\tilde{\lambda},\eta).
\end{equation}

Let us now consider the collinear limits. For a particular component
expression
$$
F_5^{tree,MHV}=\langle
\phi^{AB}_1\phi^{AB}_2g^+_3g^+_4g^+_5|T^{(0)}_{AB}(q)|0\rangle
$$
one has
\begin{eqnarray}
&& F_5^{tree,MHV}\stackrel{\phi_1^{AB} || \phi_2^{AB}} \rightarrow 0, \nonumber \\
&& F_5^{tree,MHV}\stackrel{\phi_2^{AB}||g^+_3} \rightarrow \mbox{Split}_{\phi}(\phi,g^+)F_4^{tree,MHV}, \nonumber \\
&& F_5^{tree,MHV}\stackrel{g^+_3||g^+_4} \rightarrow \mbox{Split}_{-}(g^+,g^+)F_4^{tree,MHV}, \nonumber \\
&& F_5^{tree,MHV}\stackrel{g^+_5|| \phi_1^{AB}} \rightarrow
\mbox{Split}_{\phi}(g^+,\phi) F_4^{tree,MHV},
\end{eqnarray}
where the splitting functions are defined by the following
expressions:
\begin{eqnarray}
&& \mbox{Split}_{\phi}(\phi,g^+) = \frac{1}{\langle ij \rangle}\sqrt{\frac{z}{1-z}}, \nonumber \\
&& \mbox{Split}_{\phi}(g^+,\phi) = \frac{1}{\langle ij \rangle} \sqrt{\frac{1-z}{z}}, \nonumber \\
&& \mbox{Split}_{-}(g^+,g^+) = \frac{1}{\langle ij \rangle}
\frac{1}{\sqrt{z(1-z)}},
\end{eqnarray}
where $i$ and $j$ correspond to collinear momenta.

Thus, similar to the collinear limit in the case of the MHV
amplitudes the following relation holds
$$
F_{n}^{tree,MHV}( \ldots, p_{a}^{\lambda_{i}},
p_{b}^{\lambda_{i+1}}, \ldots) \stackrel{i||i+1} \rightarrow \ \
\sum_{\lambda,c} \mbox{Split}_{-\lambda}(a^{\lambda_{i}},
b^{\lambda_{i+1}}, z) F_{n-1}^{tree, MHV}(\ldots, p_{c}^{\lambda},
\ldots).
$$

\end{document}